\documentclass[english,prd,superscriptaddress,nofootinbib,preprintnumbers]{revtex4}
\usepackage[latin1]{inputenc}
\usepackage{bm}
\usepackage{amsmath}
\usepackage{amssymb}
\usepackage{graphicx}
\usepackage{esint}

\makeatletter

\@ifundefined{textcolor}{}
{%
 \definecolor{BLACK}{gray}{0}
 \definecolor{WHITE}{gray}{1}
 \definecolor{RED}{rgb}{1,0,0}
 \definecolor{GREEN}{rgb}{0,1,0}
 \definecolor{BLUE}{rgb}{0,0,1}
 \definecolor{CYAN}{cmyk}{1,0,0,0}
 \definecolor{MAGENTA}{cmyk}{0,1,0,0}
 \definecolor{YELLOW}{cmyk}{0,0,1,0}
}

\usepackage{bm}\usepackage{color}

\usepackage{amsfonts}\usepackage{dcolumn}%\usepackage{hyperref}

\def\be{\begin{equation}}
\def\ee{\end{equation}}
\def\ba{\begin{eqnarray}}
\def\ea{\end{eqnarray}}
\def\bs{\begin{subequations}}
\def\es{\end{subequations}}

\def\Mpl{M_{\rm pl}}

\usepackage{color}

\usepackage{babel}

\makeatother

\begin{document}
\title{Planck constraints on single-field inflation}

\author{Shinji Tsujikawa}
\affiliation{Department of Physics, Faculty of Science, Tokyo University of Science, 
1-3, Kagurazaka, Shinjuku-ku, Tokyo 162-8601, Japan}

\author{Junko Ohashi}
\affiliation{Department of Physics, Faculty of Science, Tokyo University of Science, 
1-3, Kagurazaka, Shinjuku-ku, Tokyo 162-8601, Japan}

\author{Sachiko Kuroyanagi}
\affiliation{Department of Physics, Faculty of Science, Tokyo University of Science, 
1-3, Kagurazaka, Shinjuku-ku, Tokyo 162-8601, Japan}

\author{Antonio De Felice}
\affiliation{TPTP \& NEP, The Institute for Fundamental Study, Naresuan University,
Phitsanulok 65000, Thailand}
\affiliation{Thailand Center of Excellence in Physics, Ministry of Education,
Bangkok 10400, Thailand}

\begin{abstract}

We place observational constraints on slow-variation single-field 
inflationary models by carrying out the cosmological Monte Carlo 
simulation with the recent data of Planck combined 
with the WMAP large-angle polarization, baryon acoustic oscillations, 
and ACT/SPT temperature data.
Our analysis covers a wide variety of models with second-order  
equations of motion-- including potential-driven slow-roll inflation, 
non-minimally coupled models, running kinetic couplings, 
Brans-Dicke theories, potential-driven 
Galileon inflation, field-derivative couplings to the Einstein tensor, 
and k-inflation. In the presence of running kinetic exponential 
couplings, covariant Galileon terms, 
and field-derivative couplings, the tensor-to-scalar ratio of the 
self-coupling potential $V(\phi)=\lambda \phi^4/4$ gets smaller 
relative to that in standard slow-roll inflation, 
but the models lie outside the $1\sigma$ observational contour.
We also show that k-inflation models can be tightly constrained 
by adding the bounds from the scalar non-Gaussianities.
The small-field inflationary models with asymptotic flat Einstein-frame 
potentials in the regime $\phi \gg M_{\rm pl}$ generally fit the data 
very well. These include the models such as K\"{a}hler-moduli inflation, 
non-minimally coupled Higgs inflation,
and inflation in Brans-Dicke theories in the presence of the potential 
$V(\phi)=3M^2 (\phi-M_{\rm pl})^2/4$ with the Brans-Dicke parameter 
$\omega_{\rm BD} \lesssim {\cal O}(1)$
(which covers the Starobinsky's model $f(R)=R+R^2/(6M^2)$ 
as a special case). 

\end{abstract}

\date{\today}

\maketitle

%%%%%%%%%%%%%%
\section{Introduction}
\label{introsec}
%%%%%%%%%%%%%%

Inflation is an elegant idea to resolve the horizon, flatness, and 
monopole problems plagued in standard big bang 
cosmology \cite{Alexei,oldinf}.
The simplest inflationary scenario is based on a single scalar field 
(called inflaton) with a nearly flat 
potential \cite{new,chaotic,natural,hybrid} 
(see Refs.~\cite{review} for reviews).
The quantum fluctuations of inflaton can be responsible for the 
temperature anisotropies observed in the 
Cosmic Microwave Background (CMB).
The slow-roll single-field inflationary models predict nearly 
scale-invariant density perturbations \cite{oldper}, 
whose property is consistent with the CMB anisotropies measured by 
COBE \cite{COBE} and WMAP \cite{WMAP1}.

Recently, the Planck mission released the high-precision data of CMB 
temperature anisotropies up to the multipoles 
$\ell \lesssim 2500$ \cite{Adecosmo}.
The Planck data, combined with the WMAP large-angle polarization (WP)
measurement \cite{WMAP9}, showed that the spectral index $n_s$ of 
curvature perturbations 
is constrained to be $n_s=0.9603 \pm 0.0073$ at the 
wave number $k_0=0.002$\,Mpc$^{-1}$ \cite{Adeinf}.
The exact scale-invariance ($n_s=1$) is ruled out at more 
than 5$\sigma$ confidence level (CL). 
The tensor-to-scalar ratio $r$ is bounded
to be $r<0.11$ (95\,\%\,CL) at $k_0=0.002$\,Mpc$^{-1}$. 
These constraints are powerful to discriminate between a host of 
inflationary models (see Refs.~\cite{afterplanck} for observational constraints
on particular models after the data release of Planck).

The non-Gaussianities of curvature perturbations provide additional 
information to break the degeneracy between 
models \cite{Salopek,KS2001,Bartolo,Maldacena}.
In the context of single-field slow-variation inflationary models, 
the non-linear estimator $f_{\rm NL}^{\rm local}$ 
in the squeezed limit is as small as the  
orders of slow-variation parameters \cite{Maldacena,Cremi,Chen,Kaplan,DT13}.
The WMAP9 data showed that the models with purely Gaussian perturbations 
of the local shape ($f_{\rm NL}^{\rm local}=0$) are outside 
the 68\,\%\,CL observational contour \cite{WMAPfnl}.
However, the more high-precision Planck data 
constrained the non-linear estimator 
to be $f_{\rm NL}^{\rm local}=2.7 \pm 5.8$ (68\,\%\,CL) \cite{Adenon}, 
which means that single-field slow-variation inflationary 
models are consistent with the data.
The non-linear parameters of equilateral and orthogonal shapes are 
bounded to be $f_{\rm NL}^{\rm equil}=-42 \pm 75$ and 
$f_{\rm NL}^{\rm ortho}=-25 \pm 39$  (68\,\%\,CL) from 
the Planck measurement.
This information is useful to constrain models with the small
scalar propagation speed $c_s$ (such as 
k-inflation \cite{kinf,kinf2}, Galileons \cite{Nicolis,Cedric}, 
and effective field theory of inflation \cite{Weinberg,Cheung}), because 
$|f_{\rm NL}^{\rm equil}|$ and $|f_{\rm NL}^{\rm ortho}|$ can 
be much larger than 1 in those 
models \cite{Seery,Chen,Chen2,Mizuno,ATnongau0,KobaGa,Gao,DT,Senatore}.

In the light of Planck data, we put observational constraints on slow-variation 
single-field inflationary models based on the Horndeski's most general 
scalar-tensor theories \cite{Horndeski,DGSZ,Char} by running the 
Cosmological Monte-Carlo (CosmoMC) code \cite{cosmomc,Lewis}.
The Lagrangian of the Horndeski's theories is constructed to 
keep the field equations of motion up to second order for 
avoiding the Ostrogradski instability \cite{Ostro}.
This Lagrangian covers a wide variety of gravitational 
theories with one scalar degree of freedom-- such as 
standard slow-roll inflation \cite{new,chaotic,natural,hybrid}, 
non-minimally coupled models \cite{FM,Fakir,Higgs}, 
running kinetic couplings \cite{Nakayama,Reza}, 
Brans-Dicke theories \cite{Brans} 
(including $f(R)$ gravity \cite{Alexei}), 
Galileon inflation \cite{KMY,Burrage,Kamada}, 
field derivative couplings to gravity \cite{Amendola93,Germani},
and k-inflation \cite{kinf,kinf2}.
For the inflationary scenarios based on the Horndeski's theories, the power spectra 
and the non-Gaussianities of scalar and tensor perturbations 
have been already derived in Refs.~\cite{Koba2011,Gao,DT,Kobaten}.
We apply those results to concrete models of inflation mentioned above.

In order to compare theoretical predictions of $n_s$ and $r$ with observations, 
we run the CosmoMC code with the latest data of 
Planck \cite{Adecosmo}, WP \cite{WMAP9}, 
Baryon Acoustic Oscillations (BAO) \cite{BAO1,BAO2,BAO3}, 
and ACT/SPT temperature data
of high multipoles (high-$\ell$) \cite{Das,Rei}.
Since the consistency relation between the tensor-to-scalar ratio $r$
and the tensor spectral index $n_t$ is generally different depending on 
the models, we need to be careful to implement this information 
properly in the likelihood analysis. Apart from the cases
of Galileons and k-inflation, however, the consistency 
relation for the models mentioned above is the same as that 
of standard inflation ($r=-8n_t$).
In k-inflation, we study the power-law inflationary scenario based on 
the dilatonic ghost condensate \cite{kinf,Piazza} and 
Dirac-Born-Infeld (DBI) \cite{DBI,DBI2} models. 
In this case, the inflationary observables can be expressed in terms of 
the scalar propagation speed $c_s$ \cite{Ohashi2011}. 
This property is useful to place further bounds on $c_s$ from the information of 
scalar non-Gaussianities.

This paper is organized as follows.
In Sec.~\ref{Horndeskisec}, we present the background equations of 
motion in the Horndeski's theories and introduce the slow-variation 
parameters on the quasi de-Sitter background.
We also review the formulas of scalar and tensor 
power spectra as well as the scalar non-Gaussianities.
In Sec.~\ref{classification}, we classify inflationary models 
and evaluate a number of observables in each model.
In Sec.~\ref{constsec}, we carry out the likelihood analysis to 
test for each inflationary model with the latest observational data. 
Sec.~\ref{conclusions} is devoted to conclusions.

%%%%%%%%%%%%%%%%%%%%%%%%%%%%%
\section{Horndeski's theories and 
the perturbations generated during inflation}
\label{Horndeskisec}
%%%%%%%%%%%%%%%%%%%%%%%%%%%%%

We start with the action of the most general scalar-tensor theories
with second-order equation of motion \cite{Horndeski,DGSZ,Char,Koba2011}
\be
{\cal S} = \int d^{4}x \sqrt{-g}\
\biggl[\frac{M_{\rm pl}^{2}}{2}\, R+P(\phi,X)
-G_{3}(\phi,X)\,\Box\phi 
+{\cal L}_{4}+{\cal L}_{5} \biggr]\,,
\label{action}
\ee
where $g$ is the determinant of the metric tensor 
$g_{\mu\nu}$, $M_{{\rm pl}}$ is the reduced Planck mass, 
$R$ is the Ricci scalar, and 
\ba
\mathcal{L}_{4} & = & G_{4}(\phi,X)\, R+G_{4,X}\,
[(\Box\phi)^{2}-(\nabla_{\mu}\nabla_{\nu}\phi)\,(\nabla^{\mu}\nabla^{\nu}\phi)]\,,\\
\mathcal{L}_{5} & = & G_{5}(\phi,X)\, G_{\mu\nu}\,(\nabla^{\mu}\nabla^{\nu}\phi)
-\frac{1}{6}G_{5,X}[(\Box\phi)^{3}-3(\Box\phi)\,(\nabla_{\mu}\nabla_{\nu}\phi)\,
(\nabla^{\mu}\nabla^{\nu}\phi)+2(\nabla^{\mu}\nabla_{\alpha}\phi)\,
(\nabla^{\alpha}\nabla_{\beta}\phi)\,(\nabla^{\beta}\nabla_{\mu}\phi)]\,.
\ea
$P$ and $G_{i}$'s ($i=3,4,5$) are functions in terms of $\phi$
and $X=-\partial^{\mu}\phi\partial_{\mu}\phi/2$ with the partial
derivatives $G_{i,X}\equiv\partial G_{i}/\partial X$, and 
$G_{\mu\nu}=R_{\mu\nu}-g_{\mu\nu}R/2$ is the Einstein tensor 
($R_{\mu\nu}$ is the Ricci tensor). 

On the flat Friedmann-Lema\^{i}tre-Robertson-Walker background 
with the scale factor $a(t)$ ($t$ is cosmic time), 
the field equations of motion are given by \cite{Koba2011,Gao,ATnongau0,DT13}
\ba
&  & 3\Mpl^{2}H^{2}F=P_{,X}\dot{\phi}^2-P
-(G_{{3,\phi}}-12\,{H}^{2}G_{{4,X}}+9\,{H}^{2}G_{{5,\phi}} ){\dot{\phi}^{2}}
-6HG_{{4,\phi}}\dot{\phi}\nonumber \\
&  &~~~~~~~~~~~~~~~~~
-(6\, G_{{4,\phi X}}-3\, G_{{3,X}}-5\, G_{{5,X}}{H}^{2} )
H{\dot{\phi}^{3}}-3\left(G_{{5,\phi X}}-2\, G_{{4,{\it XX}}}\right)
H^{2}\dot{\phi}^{4}
+{H}^{3}G_{{5,{\it XX}}}{\dot{\phi}^{5}}\,,\label{E1d}\\
 &  & (1-4\delta_{G4X}-2\delta_{G5X}+2\delta_{G5\phi})
 \epsilon=\delta_{PX}+3\delta_{G3X}-2\delta_{G3\phi}+6\,
 \delta_{G4X}-\delta_{G4\phi}-6\,\delta_{G5\phi}+3\,\delta_{G5X}+12\,\delta_{G4XX}+2\,\delta_{G5XX}\nonumber \\
&  & ~~~~~~~~~~~~~~~~~~~~~~~~~~~~~~~~~~~~~~~~~~~~
-10\,\delta_{G4\phi X}+2\,\delta_{G4\phi\phi}-8\,\delta_{G5\phi X}
+2\,\delta_{G5\phi\phi}-\delta_{\phi}(\delta_{G3X}+4\,\delta_{G4X}-\delta_{G4\phi}\nonumber \\
&  & ~~~~~~~~~~~~~~~~~~~~~~~~~~~~~~~~~~~~~~~~~~~~
+8\,\delta_{G4XX}+3\,\delta_{G5X}-4\,\delta_{G5\phi}
+2\,\delta_{G5XX}-2\delta_{G4\phi X}-4\,\delta_{G5\phi X})\,,
\label{E1}
\ea
where $H=\dot{a}/a$ is the Hubble parameter (a dot represents a derivative with 
respect to $t$), 
\be
F=1+2G_{4}/\Mpl^{2}\,,
\ee
and 
\ba
\hspace{-0.4cm} &  & \epsilon=-\frac{\dot{H}}{H^{2}}\,,\quad\delta_{\phi}=
\frac{\ddot{\phi}}{H\dot{\phi}}\,,\quad\delta_{PX}=\frac{P_{,X}X}{\Mpl^{2}H^{2}F}\,,
\quad\delta_{G3X}=\frac{G_{3,X}\dot{\phi}X}{\Mpl^{2}HF}\,,\quad\delta_{G3\phi}
=\frac{G_{3,\phi}X}{\Mpl^{2}H^{2}F}\,,\quad\delta_{G4X}=\frac{G_{4,X}X}{\Mpl^{2}F}\,,\nonumber \\
\hspace{-0.4cm} &  & \delta_{G4\phi}=\frac{G_{4,\phi}\dot{\phi}}{\Mpl^{2}HF}\,,
\quad\delta_{G4\phi X}=\frac{G_{4,\phi X}\dot{\phi}X}{\Mpl^{2}HF}\,,\quad
\delta_{G4\phi\phi}=\frac{G_{4,\phi\phi}X}{\Mpl^{2}H^{2}F}\,,\quad
\delta_{G4XX}=\frac{G_{4,XX}X^{2}}{\Mpl^{2}F}\,,\quad
\delta_{G5\phi}=\frac{G_{5,\phi}X}{\Mpl^{2}F}\,,\nonumber \\
\hspace{-0.4cm} &  & \delta_{G5X}=\frac{G_{5,X}H\dot{\phi}X}{\Mpl^{2}F}\,,\quad
\delta_{G5XX}=\frac{G_{5,XX}H\dot{\phi}X^{2}}{\Mpl^{2}F}\,,\quad
\delta_{G5\phi X}=\frac{G_{5,\phi X}X^{2}}{\Mpl^{2}F}\,,\quad
\delta_{G5\phi\phi}=\frac{G_{5,\phi\phi}\dot{\phi}X}{\Mpl^{2}HF}\,.
\label{slowva}
\ea

In order to realize the condition $\epsilon \ll 1$, we require that all 
the parameters defined in Eq.~(\ref{slowva}) are much smaller than 1.
Taking the time derivative of the quantity $\delta_{G3 X}$, we obtain 
\be
\eta_{G3X} \equiv \frac{\dot{\delta}_{G3X}}{H \delta_{G3X}}
=\frac{2\delta_{\phi} \delta_{G3XX}}{\delta_{G3X}}
+\frac{2\delta_{G3\phi X}}{\delta_{G3X}}+
3\delta_{\phi}+\epsilon-\delta_F\,,
\label{etaG3X}
\ee
where $\delta_{G3 XX}=G_{3,XX}\dot{\phi}X^2/(\Mpl^2 HF)$, 
$\delta_{G3\phi X}=G_{3,\phi X} X^2/(\Mpl^2 H^2 F)$ 
and $\delta_F=\dot{F}/(HF)$.
Equation (\ref{etaG3X}) shows that the quantity $\delta_{G3\phi X}$ 
is second order of $\epsilon$. 
Likewise, we find 
\be
\{ \delta_{G3\phi X}, \delta_{G3\phi \phi}, \delta_{G4\phi X}, 
\delta_{G4\phi \phi}, \delta_{G5\phi X}, \delta_{G5\phi \phi} \}
={\cal O} (\epsilon^2)\,,
\ee
where 
$\delta_{G3 \phi \phi}=G_{3,\phi \phi} \dot{\phi} X/(\Mpl^2 H^3 F)$.
{}From Eq.~(\ref{E1}), it follows that 
\be
\epsilon=\delta_{PX}+3\delta_{G3X}-2\delta_{G3\phi}+6\,\delta_{G4X}
-\delta_{G4\phi}-6\,\delta_{G5\phi}+3\,\delta_{G5X}+12\,\delta_{G4XX}
+2\,\delta_{G5XX}+{\cal O} (\epsilon^2)\,.
\label{epsi}
\ee
The parameter $\delta_F$ is related to $\delta_{G4 \phi}$ via
\be
\delta_F=2\delta_{G4 \phi}+{\cal O} (\epsilon^2)\,.
\label{delF}
\ee

For later convenience, we define the number of e-foldings as
$N(t)=\ln a(t_f)/a(t)$, where $a(t)$ and $a(t_f)$ are the scale factors 
at time $t$ during inflation and at the end of inflation respectively.
Since $dN/dt=-H(t)$, it follows that 
\be
N(t)=-\int_{t_f}^t H(\tilde{t})\,d\tilde{t}\,.
\label{efold}
\ee
The field value $\phi_f$ at the end of inflation is known 
by solving $\epsilon(\phi_f)=1$.
The number of e-foldings when the perturbations relevant 
to the CMB temperature anisotropies crossed the Hubble 
radius is in the range $50<N<60$ \cite{Leach,Adeinf}.

In order to evaluate the $n$-point correlation functions of scalar and tensor 
perturbations ($n=2,3$), it is convenient to 
choose the ADM metric \cite{ADM} about the flat 
Friedmann-Lema\^{i}tre-Robertson-Walker background
\be
ds^2=-\left[ (1+\alpha)^2-a^{-2}(t) e^{-2{\cal R}} (\partial \psi)^2
\right] dt^2+2\partial_{i} \psi dt dx^i+a^2(t) \left( e^{2{\cal R}}
\delta_{ij}+h_{ij} \right) dx^i dx^j\,,
\label{permet}
\ee
where $\alpha, \psi, {\cal R}$ are scalar perturbations, and 
$h_{ij}$ is the tensor perturbation.
The uniform-field gauge ($\delta \phi=0$) is chosen to
fix the time component of a gauge-transformation vector $\xi^{\mu}$.
The scalar perturbation $E_{,ij}$ appearing in the metric (\ref{permet})
is gauged away to fix the spatial component of $\xi^{\mu}$.

The linear perturbation equations can be derived by expanding 
the action (\ref{action}) up to second order of perturbations. 
{}From the momentum and Hamiltonian constraints, 
the scalar perturbations $\alpha$ and $\psi$ are related to the curvature
perturbation ${\cal R}$.
Then, we obtain the second-order action of ${\cal R}$, 
as \cite{Koba2011,Gao,DT}
\be
{\cal S}_{s}^{(2)}=\int dt\,d^{3}x\, a^{3}Q_s \left[\dot{{\cal R}}^{2}
-\frac{c_{s}^{2}}{a^{2}}\,(\partial{\cal R})^{2}\right]\,.
\label{secondaction}
\ee
The quantities $Q_s$ and $c_s^2$ are defined by 
\ba
Q_s &\equiv& \frac{w_{1}(4w_{1}w_{3}+9w_{2}^{2})}{3w_{2}^{2}}\,,
\label{Qdef}\\
c_{s}^{2} &\equiv& \frac{3(2w_{1}^{2}w_{2}H-w_{2}^{2}w_{4}
+4w_{1}\dot{w}_{1}w_{2}-2w_{1}^{2}\dot{w}_{2})}
{w_{1}(4w_{1}w_{3}+9w_{2}^{2})}\,,
\label{cs2}
\ea
where 
\ba
w_{1} & = & \Mpl^{2}F (1-4\delta_{G4X}-2\delta_{G5X}+2\delta_{G5\phi})\,,\\
w_{2} & = & 2\Mpl^{2}HF (1-\delta_{G3X}-8 \delta_{G4X}-8\delta_{G4XX}
+\delta_{G4\phi}+2\delta_{G4 \phi X}-5 \delta_{G5X}-2\delta_{G5XX}
+6\delta_{G5\phi}+4\delta_{G5\phi X})\,,\\
w_{3} & = & -9M_{{\rm pl}}^{2}H^{2}F (1-\delta_{PX}/3-2\delta_{PXX}/3
-4\delta_{G3X}-2\delta_{G3XX}+2\delta_{G3\phi}/3+2\delta_{G3 \phi X}/3
-14 \delta_{G4X}-32 \delta_{G4XX} -8 \delta_{G4XXX} \nonumber \\
& &
+2\delta_{G4\phi}+10\delta_{G4\phi X}
+4\delta_{G4\phi XX}-10\delta_{G5X}-26\delta_{G5XX}/3-4\delta_{G5XXX}/3
+12\delta_{G5\phi}+18\delta_{G5\phi X}+4\delta_{G5 \phi XX})\,, \\
w_{4} & = & \Mpl^{2}F (1-2\delta_{G5\phi}-2\delta_{G5X}\delta_{\phi} )\,,
\ea
and
\ba
& &
\delta_{PXX}=\frac{X^2 P_{,XX}}{\Mpl^2 H^2 F}\,,\quad
\delta_{G4XXX}=\frac{G_{4,XXX}X^3}{\Mpl^2 F}\,,\quad
\delta_{G4\phi XX}=\frac{G_{4,\phi XX}\dot{\phi}X^2}{\Mpl^2 H F}\,,
\nonumber \\
& &
\delta_{G5 XXX}=\frac{G_{5,XXX}H\dot{\phi}X^3}{\Mpl^2 F}\,,\quad
\delta_{G5\phi XX}=\frac{G_{5,\phi XX}X^3}{\Mpl^2 F}\,.
\ea
The terms $\delta_{G4\phi XX}$ and $\delta_{G5\phi XX}$
are of the order of $\epsilon^2$.

At leading order in slow-variation parameters, 
we have
\ba
Q_s &=& \Mpl^2 F q_s\,,\\
q_s &\equiv&
\delta_{PX}+2\delta_{PXX}+6\delta_{G3X}+6\delta_{G3XX}
+6\delta_{G4X}+48 \delta_{G4XX}+24\delta_{G4XXX}  \nonumber \\
& &
+6\delta_{G5X}+14\delta_{G5XX}+4\delta_{G5XXX}
-2\delta_{G3\phi}-6\delta_{G5\phi}\,,\label{qs} \\
\epsilon_s &\equiv& \frac{Q_s c_s^2}{ \Mpl^2 F}
=\delta_{PX}+4\delta_{G3X}+6\delta_{G4X}+20\delta_{G4XX}
+4\delta_{G5X}+4\delta_{G5XX}-2\delta_{G3\phi}-6\delta_{G5\phi}\,,
\label{eps}
\ea
by which the scalar propagation speed squared 
can be expressed as
\be
c_s^2=\frac{\epsilon_s}{q_s}\,.
\label{cs2d}
\ee
The power spectrum of curvature perturbations 
is given by \cite{Koba2011,Gao,DT} 
\be
{\cal P}_{{\cal R}}=\frac{H^{2}}{8\pi^{2}
M_{\rm pl}^{2}\epsilon_{s}Fc_{s}} \bigg|_{c_sk=aH}\,,
\label{scalarpower}
\ee
which should be evaluated at the epoch 
when the mode with a wave number $k$ crossed $c_sk=aH$ during inflation.
The scalar spectral index reads
\be
n_s-1 \equiv \frac{d\ln{\cal P}_{{\cal R}}}{d\ln k}
\bigg|_{c_{s}k=aH}
=-2\epsilon-\eta_{s}-\delta_F-s\,,
\label{nR}
\ee
where 
\be
\eta_s \equiv \frac{\dot{\epsilon}_s}{H \epsilon_s}\,,\qquad
s \equiv \frac{\dot{c}_s}{Hc_s}\,.
\ee
The running spectral index is defined by 
\be
\alpha_s \equiv \frac{d n_s}{d \ln k} \bigg|_{c_{s}k=aH}\,,
\ee
which is of the order of $\epsilon^2$ from Eq.~(\ref{nR}).

The transverse and traceless tensor perturbation $h_{ij}$ 
can be decomposed 
into two independent polarization modes, as 
$h_{ij}=h_{+}e_{ij}^{+}+h_{\times} e_{ij}^{\times}$.
The tensors $e_{ij}^{\lambda}$ (where $\lambda=+,\times$)
satisfy the relations 
$e_{ij}^{+} ({\bm k}) e_{ij}^{+} (-{\bm k})^*=2$,
$e_{ij}^{\times} ({\bm k}) e_{ij}^{\times} (-{\bm k})^*=2$,
and $e_{ij}^{+} ({\bm k}) e_{ij}^{\times} (-{\bm k})^*=0$
in Fourier space.
The second-order action for the tensor perturbation 
is \cite{Koba2011,Gao,DT} 
\begin{equation}
{\cal S}_{t}^{(2)}=\sum_{\lambda=+,\times}\int dt\, d^{3}x\, a^{3} 
Q_{t}\left[ \dot{h}_{\lambda}^{2}
-\frac{c_{t}^2}{a^2} (\partial h_{\lambda})^2 \right]\,,
\label{ST}
\end{equation}
where 
\begin{eqnarray}
Q_{t} &=&\frac{1}{4}w_{1}=\frac{1}{4} M_{{\rm pl}}^{2}F
(1-4\delta_{G4X}-2\delta_{G5X}+2\delta_{G5\phi})\,,
\label{QT} \\ 
c_{t}^{2} &=& \frac{w_{4}}{w_{1}}=
1+4\delta_{G4X}+2\delta_{G5X}-4\delta_{G5\phi}
+{\cal O} (\epsilon^2)\,.
\label{cT}
\end{eqnarray}
The tensor power spectrum reads
\be
{\cal P}_{h}=\frac{H^{2}}
{2\pi^{2}Q_t c_t^3} \biggl|_{c_tk=aH} \simeq 
\frac{2H^2}{\pi^2 M_{\rm pl}^2 F} \biggl|_{k=aH} \,,
\label{Ph}
\ee
where, in the second approximate equality, we have taken
leading-order terms.

When both ${\cal R}$ and $h_{\lambda}$ approach approximately constant
values during inflation, the tensor-to-scalar ratio 
can be evaluated as 
\be
r=\frac{{\cal P}_h}{{\cal P}_{\cal R}}
\simeq 16 c_s \epsilon_s\,.
\label{ratio}
\ee
We define the tensor spectral index and its running, as 
\ba
n_t & \equiv & \frac{d\ln{\cal P}_h}{d\ln k}\bigg|_{k=aH}
=-2\epsilon-\delta_F\,,\label{nt} \\
\alpha_t &\equiv& \frac{d n_t}{d \ln k} \bigg|_{k=aH}\,,
\ea
where $\alpha_t$ is of the order of $\epsilon^2$.
Using Eqs.~(\ref{epsi}), (\ref{delF}), and (\ref{eps}), we obtain 
the following consistency relation 
\be
r=-8c_s \left( n_t-2\delta_{G3X}-16\delta_{G4XX}
-2\delta_{G5X}-4\delta_{G5XX} \right)\,. 
\label{consistency}
\ee

In order to avoid ghosts and Laplacian instabilities, 
we require the conditions $Q_s>0$, $c_s^2>0$, 
$Q_t>0$, and $c_t^2>0$, i.e., 
\be
F>0\,,\qquad q_s>0\,,\qquad \epsilon_s>0\,.
\ee
We focus on the models in which these conditions 
are satisfied.

The non-Gaussianities of curvature perturbations generated in the 
Horndeski's theories were evaluated in Refs.~\cite{Gao,DT,DT13}.
The bispectrum ${\cal A}_{\cal R}$ is related to the three-point 
correlation function of ${\cal R}$, as
\be
\langle {\cal R} ({\bm k}_1) {\cal R} ({\bm k}_2) {\cal R} ({\bm k}_3)
\rangle
=(2\pi)^7 \delta^{(3)} ({\bm k}_1+{\bm k}_2+{\bm k}_3)
({\cal P}_{\cal R})^2 \frac{{\cal A}_{\cal R} (k_1,k_2,k_3)}
{\prod_{i=1}^3 k_i^3}\,.
\ee
We also define the non-linear estimator $f_{\rm NL}$, as
\be
f_{\rm NL}=\frac{10}{3} \frac{{\cal A}_{\cal R}}
{\sum_{i=1}^3 k_i^3}\,.
\label{fnldef}
\ee
The leading-order bispectrum was derived in Refs.~\cite{Gao,DT13} 
on the de Sitter background.
In Ref.~\cite{DT13}, the authors computed the three-point correlation 
function by taking into account all the possible slow-variation corrections
to the leading-order term.
In this case, the resulting bispectrum is valid for any shape 
of non-Gaussianities with the momentum triangle satisfying 
${\bm k}_1+{\bm k}_2+{\bm k}_3=0$.
Under the slow-variation approximation used above,
the non-linear estimator $f_{\rm NL}^{\rm local}$ 
in the squeezed limit ($k_3 \to 0$, $k_1 \to k_2$) 
is given by \cite{DT13}
\be
f_{{\rm NL}}^{{\rm local}}=\frac{5}{12} (1-n_s)\,.
\label{flocal}
\ee
Since $f_{\rm NL}^{\rm local}$ is of the order of $\epsilon$, 
the Planck bound $f_{\rm NL}^{\rm local}=2.7 \pm 5.8$ (68\,\%\,CL) is satisfied 
for all the slow-variation single-field models based on the Horndeski's theories. 
There are some non slow-roll inflationary models in which the relation 
(\ref{flocal}) can be violated \cite{nonslow}, but we generally 
require the tunings of model parameters and initial conditions
to satisfy the constraints of $n_s$ and $r$.
Hence we do not consider such specific models in our paper.

In Ref.~\cite{DT13}, it was shown that the leading-order bispectrum
(of the order of $\epsilon^0$) can be expressed by the linear combination 
of two bases $S_7^{\rm equil}$ and $S_7^{\rm ortho}$, as
\be
{\cal A}_{\cal R}^{\rm lead}=c_1 S_7^{\rm equil}+
c_2 S_7^{\rm ortho}\,.
\label{bispectrum}
\ee
The coefficients $c_1$ and $c_2$ are
\ba
\hspace{-0.5cm}c_1 &=& \frac{13}{12}\biggl[\frac{1}{24}\left(1-\frac{1}{c_{s}^{2}}\right)
\left(2+3\beta\right)+\frac{\mu}{12\Sigma}\left(2-3\beta\right) \nonumber \\
\hspace{-0.5cm}& &~~~~~-\frac{2-3\beta}{6} 
\frac{\delta_{G3X}+\delta_{G3XX} +4(3\delta_{G4XX}+2\delta_{G4XXX})
+\delta_{G5X}+5\delta_{G5XX}+2\delta_{G5XXX}}{\epsilon_{s}}
\nonumber \\
\hspace{-0.5cm}& &~~~~~
+\frac{\delta_{G3X}+6\delta_{G4XX}+\delta_{G5X}+\delta_{G5XX}}{3\epsilon_{s}c_{s}^{2}}\biggr]\,,\\
\hspace{-0.5cm}c_2 &=& \frac{14-13\beta}{12}\left[\frac{1}{8}\left(1-\frac{1}{c_{s}^{2}}\right)
-\frac{\mu}{4\Sigma}
+\frac{\delta_{G3X}+\delta_{G3XX} +4(3\delta_{G4XX}+2\delta_{G4XXX})
+\delta_{G5X}+5\delta_{G5XX}+2\delta_{G5XXX}}{2\epsilon_{s}}\right],
\ea
where $\beta=1.1967996 \cdots$, and
\ba
\mu & = & \frac{F^{2}}{3}[3X^{2}P_{,XX}+2X^{3}P_{,XXX}+
3H\dot{\phi}(XG_{3,X}+5X^{2}G_{3,XX}+2X^{3}G_{3,XXX})-2(2X^{2}G_{3,\phi X}
+X^{3}G_{3,\phi XX})\nonumber \\
 &  & +6H^{2}(9X^{2}G_{4,XX}+16X^{3}G_{4,XXX}+4X^{4}G_{4,XXXX})
 -3H\dot{\phi}(3XG_{4,\phi X}+12X^{2}G_{4,\phi XX}+4X^{3}G_{4,\phi XXX})\nonumber \\
 &  & +H^{3}\dot{\phi}(3XG_{5,X}+27X^{2}G_{5,XX}+24X^{3}G_{5,XXX}+4X^{4}G_{5,XXXX})\nonumber \\
&  & -6H^{2}(6X^{2}G_{5,\phi X}+9X^{3}G_{5,\phi XX}+2X^{4}G_{5,\phi XXX})]\,,
\label{mudef} \\
\Sigma & = & \frac{w_{1}(4w_{1}w_{3}+9w_{2}^{2})}{12M_{{\rm pl}}^{4}}\,.
\label{Sigdef}
\ea
The shape functions $S_7^{\rm equil}$ and $S_7^{\rm ortho}$, 
which have high correlations with the equilateral and orthogonal 
templates respectively, are given by 
\ba
S_{7}^{\rm equil}=-\frac{12}{13}
\frac{1}{K}\left(1+\frac{1}{K^{2}}\,\sum_{i>j}k_{i}k_{j}+\frac{3k_{1}k_{2}k_{3}}{K^{3}}\right)\left[\frac{3}{4}\,\sum_{i}k_{i}^{4}-\frac{3}{2}\sum_{i>j}k_{i}^{2}k_{j}^{2}\right]\,,\\
S_{7}^{\rm ortho}=\frac{12}{14-13\beta} \left[-\frac{13}{12}\beta S_{7}^{\rm equil}
+\frac{4}{K} \sum_{i>j} k_i^2 k_j^2-\frac{2}{K^2} \sum_{i \neq j}
k_i^2 k_j^3-\frac12 \sum_{i}k_i^3 \right]\,,
\ea
where $K=k_1+k_2+k_3$.
These functions are normalized as $S_{7}^{\rm equil}=S_{7}^{\rm ortho}=k^3$
at $k_1=k_2=k_3 \equiv k$.
In the limit of the equilateral triangle ($k_1=k_2=k_3$), 
the leading-order non-linear parameter
$f_{\rm NL}^{\rm eq}=10{\cal A}_{\cal R}^{\rm lead}/(9k^3)$ reads
\ba
f_{{\rm NL}}^{{\rm eq}} &=&  \frac{85}{324}\left(1-\frac{1}{c_{s}^{2}}\right)-\frac{10}{81}\frac{\mu}{\Sigma}+\frac{20}{81\epsilon_{s}}
[\delta_{G3X}+\delta_{G3XX} +4(3\delta_{G4XX}+2\delta_{G4XXX})
+\delta_{G5X}+5\delta_{G5XX}+2\delta_{G5XXX}] \nonumber \\
&  & +\frac{65}{162c_{s}^{2}\epsilon_{s}}(\delta_{G3X}+6\delta_{G4XX}+\delta_{G5X}+\delta_{G5XX})\,.\label{fnleq}
\ea
For the models in which $c_s^2$ is of the order of 1, $|f_{{\rm NL}}^{{\rm eq}}|$
is at most of the order of 1. 
However, the models with $c_s^2 \ll 1$ (such as k-inflation)
are subject to be constrained from the non-Gaussianities.
 
%%%%%%%%%%%%%%%%%%%%%%
\section{Classification of single-field models}
\label{classification}
%%%%%%%%%%%%%%%%%%%%%%

In this section, we classify single-field inflationary models based on 
the Horndeski's theories and evaluate the inflationary observables 
in each model. 
We stress that our analysis is based on the slow-variation 
approximation, under which all the parameters defined in Eq.~(\ref{slowva})
are much smaller than 1. There are some specific cases in which 
the slow-variation approximation is violated, but if this occurs
for the perturbations relevant to CMB anisotropies, the scalar 
spectral index tends to be at odds with observations.
Hence we focus on the slow-variation inflationary scenario where 
the condition $|\epsilon| \ll 1$ is satisfied in the whole regime
characterized by $N<70$.

\subsection{Potential-driven slow-roll inflation}

The standard slow-roll inflation driven by a potential 
energy $V(\phi)$ of a canonical field $\phi$ is given by 
\be
P(\phi, X)=X-V(\phi),\quad
G_3=0\,,\quad G_4=0\,,\quad G_5=0\,.
\label{standinf}
\ee
In this case, $\epsilon=\epsilon_s=\delta_{PX}=\dot{\phi}^2/(2M_{\rm pl}^2 H^2)$, 
$c_s^2=1$, $s=0$, and $\delta_F=0$.
Under the slow-roll approximation ($\dot{\phi}^2/2 \ll V$),
we have $3\Mpl^2 H^2 \simeq V$ from Eq.~(\ref{E1d}).
Taking the time-derivative of this equation and using Eq.~(\ref{E1}), 
it follows that $\dot{\phi} \simeq -V_{,\phi}/(3H)$.
Then, the number of e-foldings (\ref{efold}) reads
\be
N \simeq \frac{1}{M_{\rm pl}^2} \int_{\phi_f}^{\phi}
\frac{V}{V_{,\tilde{\phi}}} d \tilde{\phi}\,.
\label{efoldstan}
\ee

The slow-roll parameters $\epsilon_s$ and $\eta_s$
reduce to $\epsilon_s \simeq \epsilon_V$ and 
$\eta_s \simeq 4\epsilon_V-2\eta_V$, where
\be
\epsilon_V \equiv \frac{\Mpl^2}{2} 
\left( \frac{V_{,\phi}}{V} \right)^2\,,
\qquad
\eta_V \equiv \frac{\Mpl^2 V_{,\phi \phi}}{V}\,.
\ee
The observables (\ref{nR}), (\ref{ratio}), and (\ref{nt}) 
reduce to
\be
n_s=1-6\epsilon_V+2\eta_V\,,\qquad 
r=-8n_t\,,\qquad
n_t=-2\epsilon_V\,.
\label{nsr}
\ee
Using Eqs.~(\ref{efoldstan})-(\ref{nsr}), we obtain the relation 
$(d\phi/dN)^2=(M_{\rm pl}^2/8)r$.
Assuming that $r$ is nearly constant, the variation $\Delta \phi$ 
of the field during inflation (corresponding to $N \approx 60$) 
can be estimated as
\be
\Delta \phi/M_{\rm pl} \approx {\cal O}(1) \times (r/0.01)^{1/2}\,.
\label{Lyth}
\ee
This is known as the Lyth bound \cite{Lythbound}, 
which relates $\Delta \phi$ with the tensor-to-scalar 
ratio $r$. 
The models with $\Delta \phi \gtrsim M_{\rm pl}$ and 
$\Delta \phi \lesssim M_{\rm pl}$
are called the ``large-field'' and ``small-field'' models, respectively.

Let us consider the power-law potential \cite{chaotic}
\be
V(\phi)=\lambda \phi^n/n\,,
\label{powerlaw}
\ee
where $n$ and $\lambda$ are positive constants.
In this case,
we have $\epsilon_V=n^2 M_{\rm pl}^2/(2\phi^2)$ and 
$\eta_V=n(n-1)M_{\rm pl}^2/\phi^2$.
The field value $\phi_f$ at the end of inflation can be
derived from the condition $\epsilon_V (\phi_f)=1$, 
that is, $\phi_f=nM_{\rm pl}/\sqrt{2}$.
{}From Eq.~(\ref{efoldstan}) the number of 
e-foldings $N$ is related to the field $\phi$, as
$\phi^2(N) \simeq 2n \left( N+n/4 \right)M_{\rm pl}^2$. 
Then, it follows that 
\be
n_{s}=1-\frac{2(n+2)}{4N+n}\,,\qquad
r=\frac{16n}{4N+n}=\frac{8n}{n+2} (1-n_s)\,.
\label{nsrstan}
\ee

For the exponential potential 
$V(\phi)=V_0 e^{-\gamma \phi/M_{\rm pl}}$ \cite{Mata,Liddle},
we have $n_s=1-\gamma^2$ and $r=8\gamma^2$, 
so this model is on the line
\be
r=8(1-n_s)\,,
\label{exp}
\ee
which corresponds to the limit $n \to \infty$ in 
the last relation of Eq.~(\ref{nsrstan}). 
Since inflation does not end for the exponential potential, 
we require the modification of the potential around 
the end of inflation.

\subsection{Non-minimally coupled models}

We proceed to non-minimally coupled theories described 
by the action  
\begin{equation}
S=\int d^{4}x\sqrt{-g}\left[\frac{M_{{\rm pl}}^{2}}{2}F(\phi)R
+\omega(\phi)X-V(\phi)\right]\,,
\label{actionsta}
\end{equation}
where $F(\phi)$, $\omega(\phi)$, and $V(\phi)$ are 
functions of $\phi$.
{}From Eqs.~(\ref{qs})-(\ref{cs2d}) and (\ref{mudef})-(\ref{Sigdef}),
we have $c_s^2=1$, $s=0$, and $\mu/\Sigma=0$. 

Under the conformal transformation  
$\hat{g}_{\mu\nu}=F(\phi)g_{\mu\nu}$, 
the action (\ref{actionsta}) recasts to the one with 
a minimally coupled scalar field (the Einstein frame) \cite{Maeda}. 
The transformed action is given by 
\begin{equation}
S_{E}=\int d^{4}x\sqrt{-\hat{g}}\left[\frac{1}{2}\Mpl^{2}\hat{R}
-\frac{1}{2}\hat{g}^{\mu\nu}\partial_{\mu}\chi\partial_{\nu}\chi
-U(\chi)\right]\,,
\label{Eaction}
\end{equation}
where a hat represents the quantities in the Einstein frame, and
\begin{equation}
U=\frac{V}{F^{2}}\,,\qquad\chi\equiv\int B(\phi)\, d\phi\,,\qquad 
B(\phi)\equiv\sqrt{\frac{3}{2}\left(\frac{\Mpl F_{,\phi}}{F}\right)^{2}
+\frac{\omega}{F}}\,.
\end{equation}
In Refs.~\cite{nonminimalper,Komatsuper,Gumjudpai}, 
it was shown that inflationary observables 
such as $n_s$ and $r$ are unchanged even after the conformal 
transformation (i.e., $\hat{n}_s=n_s$ and $\hat{r}=r$).
Then, we just need to use the formulas (\ref{nsr}) by 
replacing $\epsilon_V$ and $\eta_V$ for 
$\epsilon_U=(\Mpl^2/2)(U_{,\chi}/U)^2$ and 
$\eta_U=\Mpl^2 U_{,\chi \chi}/U$, respectively.

Let us study the non-minimal coupling models 
given by \cite{FM,Fakir,Dehnen,Higgs} 
\begin{equation}
F(\phi)=1-\xi \phi^2/\Mpl^2\,,
\end{equation}
with the non-canonical kinetic term $\omega(\phi) X$.
For the self-coupling potential $V(\phi)=\lambda \phi^4/4$, 
the presence of the non-minimal coupling allows a possibility of 
reducing $r$ \cite{Komatsuper,Gumjudpai}.
In particular, the Higgs potential $V(\phi)=(\lambda/4)(\phi^2-\phi_0^2)^2$ 
with $\lambda \sim 0.1$ and $\phi_0 \ll M_{\rm pl}$ can be 
accommodated for largely negative values of $\xi$ \cite{Higgs}.
This comes from the fact that 
the amplitude of curvature perturbations
is proportional to $\lambda/\xi^2$ in the regime $|\xi| \gg 1$.

In the Einstein frame, the power-law potential (\ref{powerlaw}) 
takes the form
\begin{equation}
U=\frac{\lambda \phi^n}{n(1-\xi x^2)^2}\,,\quad
{\rm where} \quad
x \equiv \frac{\phi}{\Mpl}\,.
\label{Uein}
\end{equation}
For $n=4$ and $\xi<0$, this potential is asymptotically flat in 
the regime $\phi \gg \Mpl$.
{}From Eq.~(\ref{nsr}), the scalar spectral index and the 
tensor-to-scalar ratio for the potential (\ref{Uein}) 
read \cite{Reza} 
\begin{eqnarray}
n_{s}-1 &=& -\frac{1}{[\omega+(6\xi-\omega)\xi x^{2}]^{2}x^{2}}\biggl\{(n-4)^{2}(6\xi-\omega)(\xi x^{2})^{3}+(24\omega-14n\omega+3n^{2}\omega+24n\xi-12n^{2}\xi)(\xi x^{2})^{2}\nonumber \\
 &  & +(-8\omega+4n\omega-3n^{2}\omega+24n\xi+6n^{2}\xi)\xi x^{2}+
 n\omega(n+2)-\mu\omega x(1-\xi x^{2})^{2}[(n-4)\xi x^{2}-n]
 \biggr\}\,,\label{nRex2}\\
r &=& -8n_{t}=
\frac{8[n+(4-n)\xi x^{2}]^{2}}{x^{2}[\omega+(6\xi-\omega)\xi x^{2}]}\,,\label{rex2}
\end{eqnarray}
where $\mu \equiv \Mpl\omega_{,\phi}/\omega$.
The scalar power spectrum (\ref{scalarpower}) is given by 
\begin{equation}
{\cal P}_{\cal R} = \frac{U^3}{12\pi^2 \Mpl^6 U_{,\chi}^2}
=\frac{\lambda M_{\rm pl}^{n-4}}{12 \pi^2 n} \frac{x^{n+2} 
[6 \xi^2 x^2+\omega
(1-\xi x^2)]}{(1-\xi x^2)^2 [n+(4-n)\xi x^2]^2}\,.
\label{Psam}
\end{equation}
For the models with constant $\omega$, we have $\mu=0$.

The Hubble parameters in the Jordan and Einstein 
frames ($H$ and $\hat{H}$ respectively) have the relation 
$\hat{H}=[H+\dot{F}/(2F)]/\sqrt{F}$. 
The number of e-foldings in the Einstein frame should be the same as 
that in the Jordan frame by properly choosing a reference 
length scale \cite{Catena}. Under the slow-roll approximation, 
the frame-independent quantity (\ref{efold}) reads \cite{Reza} 
\begin{equation}
N=\int_{\chi_f}^{\chi} \frac{U}{M_{\rm pl}^2 U_{,\chi}}d \chi
+\frac12 \ln \frac{F}{F_f}\,,
\label{efoldap}
\end{equation}
where the subscript ``$f$'' represents the value at the 
end of inflation (which is determined by the condition 
$\epsilon_U=1$).
For the potential (\ref{Uein}) it follows that 
\begin{eqnarray}
&  & N=-\frac{1}{4\xi} \ln \left|\frac{(n-4)\xi x_{f}^{2}-n}
{(n-4) \xi x^{2}-n} \right|^{\frac{3n\xi-2\omega}{n-4}}
-\frac{1}{4}\ln\left|\frac{1-\xi x^{2}}{1-\xi x_{f}^{2}}\right|
\qquad(n\neq4)\,,\label{Nap0}\\
&  & N=\frac{\omega-6 \xi}{8}(x^{2}-x_{f}^{2})
-\frac{1}{4}\ln\left|\frac{1-\xi x^{2}}{1-\xi x_{f}^{2}}\right|
\qquad(n=4)\,,\label{Nap}
\end{eqnarray}
where 
\begin{equation}
x_{f}^{2}=\frac{\omega-\xi n(4-n)-\sqrt{(\omega-2n \xi)
(\omega-6n \xi)}}{\xi [\xi (4-n)^{2}+2(\omega-6\xi)]}\,.
\label{psif}
\end{equation}
\subsection{Running kinetic couplings}

Running kinetic inflation is characterized by the action 
(\ref{actionsta}) in the presence of the field-dependent 
coupling $\omega(\phi)$ with $F(\phi)=1$.
We focus on the case of the exponential 
coupling \cite{Reza}
\begin{equation}
\omega(\phi)=e^{\mu \phi/M_{\rm pl}}\,,
\end{equation}
where $\mu$ is constant.
This is motivated by the dilatonic coupling in 
low-energy effective string theory \cite{Gas}. 
For concreteness, we take the power-law 
potential (\ref{powerlaw}). 
At the potential minimum ($\phi=0$), the coupling 
$\omega(\phi)$ is equivalent to 1.
 
The spectral index (\ref{nRex2}) and the tensor-to-scalar 
ratio (\ref{rex2}) read
\begin{equation}
n_s-1=-\frac{n}{x^2 e^{\mu x}} (n+2+\mu x)\,,\qquad
r=-8n_t=\frac{8n^2}{x^2 e^{\mu x}}\,,
\label{runningns}
\end{equation}
where $x=\phi/M_{\rm pl}$.
{}From Eq.~(\ref{efoldap}), the number of e-foldings is
\begin{equation}
N=\frac{1}{n \mu^2} \left[ (\mu x-1) e^{\mu x}-
(\mu x_f-1) e^{\mu x_f} \right]\,,
\end{equation}
where the field value at the end of inflation is known 
by solving $x_f^2 e^{\mu x_f}=n^2/2$.

\subsection{Brans-Dicke theories (including $f(R)$ gravity)}

The Brans-Dicke (BD) theory is described by the action 
\begin{equation}
S=\int d^{4}x\sqrt{-g}\left[\frac{1}{2}\Mpl \phi R
+\frac{\Mpl}{\phi}\omega_{{\rm BD}}X-V(\phi)\right]\,,
\label{Jframe}
\end{equation}
where $\omega_{{\rm BD}}$ is the BD parameter. 
Contrary to the original BD theory \cite{Brans}, 
we introduced the field potential $V(\phi)$.
Since this theory belongs to a class of the action (\ref{actionsta}), 
the action in the Einstein frame is given 
by Eq.~(\ref{Eaction}) with
\begin{equation}
U=e^{-2\gamma \chi/M_{\rm pl}}V\,,\qquad
F=\phi/M_{\rm pl}=e^{\gamma \chi/M_{\rm pl}}\,,\qquad
\gamma=1/\sqrt{3/2+\omega_{\rm BD}}\,.
\end{equation}
The scalar and tensor ghosts are absent under the 
conditions $\omega_{\rm BD}>-3/2$ and $F>0$.

The $f(R)$ theory characterized by the action 
$S=\int d^4 x \sqrt{-g}\,M_{\rm pl}^2\,f(R)/2$
is a subclass of the BD theory (\ref{Jframe}) 
with the correspondence
\begin{equation}
\frac{\phi}{M_{\rm pl}}=\frac{\partial f}{\partial R}\,,\qquad
V(\phi)=\frac{M_{\rm pl}^2}{2} \left( R \frac{\partial f}{\partial R}
-f \right)\,,\qquad \omega_{\rm BD}=0\,.
\label{fRcorre}
\end{equation}
The Starobinsky's model of inflation \cite{Alexei} 
corresponds to the Lagrangian 
\begin{equation}
f(R)=R+\frac{R^2}{6M^2}\,,
\label{fRsta}
\end{equation}
where $M$ is a constant having a dimension of mass.
In this case, the potential $V(\phi)$ 
in the Jordan frame reads
\begin{equation}
V(\phi)=\frac{3M^2}{4} \left( \phi-M_{\rm pl} \right)^2\,,
\label{poBD}
\end{equation}
where $\phi/M_{\rm pl}=1+R/(3M^2)$.

We study the BD theory described by the 
action (\ref{Jframe}) with the field potential (\ref{poBD}). 
This analysis covers the $f(R)$ model (\ref{fRsta})
as a special case ($\omega_{\rm BD}=0$).
The potential in the Einstein frame reads
\begin{equation}
U(\chi)=\frac34 M^2 M_{\rm pl}^2 \left( 
1-e^{-\gamma \chi/M_{\rm pl}} \right)^2\,.
\label{Uchi}
\end{equation}
Inflation occurs in the regime $\gamma \chi/M_{\rm pl} \gg 1$, 
which is followed by the reheating stage characterized
by the potential $U(\chi) \simeq (3/4)\gamma^2 M^2 \chi^2$.
For the potential (\ref{Uchi}), the inflationary observables 
are \cite{Reza}
\begin{equation}
n_s-1=-\frac{4 \gamma^2 (F+1)}{(F-1)^2}\,,\qquad
r=-8n_t=\frac{32 \gamma^2}{(F-1)^2}\,.
\label{Bransns}
\end{equation}
The number of e-foldings (\ref{efoldap}) yields
\begin{equation}
N=\frac{1}{2\gamma^2} (F-F_f)+\frac12 \left( 
1-\frac{1}{\gamma^2} \right) \ln \left( \frac{F}{F_f}
\right)\,,
\end{equation}
where $F_f=1+\sqrt{2} \gamma$.

\subsection{Potential-driven Galileon inflation}

The potential-driven inflation with covariant Galileon terms \cite{Kamada,Popa,Ohashi} 
belongs to a class of the action (\ref{action}) with the choice
\begin{equation}
P=X-V(\phi)\,,\qquad 
G_3=\frac{c_3}{M^3}X\,,\qquad
G_4=-\frac{c_4}{M^6}X^2\,,\qquad
G_5=\frac{3c_5}{M^9}X^2\,.
\label{Galilag}
\end{equation}
{}From Eq.~(\ref{epsi}), we have 
\begin{equation}
\epsilon=(1+{\cal A})\delta_{PX}\,,\quad
{\rm where} \quad
{\cal A} \equiv \frac{3\delta_{G3X}+18\delta_{G4X}+
5\delta_{G5X}}{\delta_{PX}}\,.
\label{epGa}
\end{equation}
Since inflation is mainly driven by the potential energy, 
Eq.~(\ref{E1d}) is approximately given by 
\begin{equation}
3M_{\rm pl}^2 H^2 \simeq V\,.
\label{Hubble1}
\end{equation}
Taking the time derivative of Eq.~(\ref{Hubble1}) and 
using Eq.~(\ref{epGa}), it follows that 
\begin{equation}
3H \dot{\phi} (1+{\cal A}) \simeq -V_{,\phi}\,.
\label{field1}
\end{equation}
{}From Eqs.~(\ref{Hubble1}) and (\ref{field1}), we have 
$\delta_{PX} \simeq \epsilon_V/(1+{\cal A})^2$ and 
hence $\epsilon \simeq \epsilon_V/(1+{\cal A})$.
The field value at the end of inflation is known by 
$\epsilon (\phi_f)=1$, i.e., 
\begin{equation}
\epsilon_V (\phi_f)=1+{\cal A} (\phi_f)\,.
\label{pend}
\end{equation}
Using Eqs.~(\ref{Hubble1}) and (\ref{field1}), 
the number of e-foldings (\ref{efold}) reads
\begin{equation}
N \simeq \frac{1}{M_{\rm pl}^2} \int_{\phi_f}^{\phi}
(1+{\cal A}) \frac{V}{V_{,\tilde{\phi}}} d\tilde{\phi}\,.
\label{Nga}
\end{equation}
In the regime ${\cal A} \gg 1$, the Galileon self-interaction 
dominates over the standard kinetic term $X$.

The quantities $q_s$ and $c_s^2$ in Eqs.~(\ref{qs}) and 
(\ref{cs2d}) reduce to 
\begin{eqnarray}
q_s &=& \delta_{PX}+6 \delta_{G3X}+54 \delta_{G4X}+
20\delta_{G5X}\,,\\
c_s^2 &=& \frac{\delta_{PX}+4 \delta_{G3X}+26 \delta_{G4X}+
8\delta_{G5X}}
{\delta_{PX}+6 \delta_{G3X}+54 \delta_{G4X}+
20\delta_{G5X}}\,.
\label{csGa}
\end{eqnarray}
In order to avoid the appearance of scalar ghosts and Laplacian instabilities
in the regime ${\cal A} \gg 1$, we demand the conditions 
$c_3 \dot{\phi}>0$, $c_4<0$, and $c_5 \dot{\phi}>0$.
In the case where either of $\delta_{G3X}$, $\delta_{G4X}$, 
$\delta_{G5X}$ dominates over $\delta_{PX}$ during inflation, 
the scalar propagation speed squared is 
$c_s^2=2/3, 13/27, 2/5$, respectively. 
This leads to the modification of the consistency 
relation $r=-8n_t$ in standard potential-driven slow-roll inflation.
Using Eqs.~(\ref{eps}), (\ref{epGa}), and (\ref{csGa}), 
the tensor-to-scalar ratio (\ref{ratio}) and the tensor spectral 
index (\ref{nt}) reduce to
\begin{eqnarray}
r &=& 16 \frac{(\delta_{PX}+4\delta_{G3X}+26 \delta_{G4X}+
8\delta_{G5X})^{3/2}}{(\delta_{PX}+6\delta_{G3X}+54\delta_{G4X}+
20\delta_{G5X})^{1/2}}\,,\label{rga}\\
n_t &=& -2(\delta_{PX}+3\delta_{G3X}+18\delta_{G4X}
+5\delta_{G5X})\,.
\end{eqnarray}
If either of $\delta_{G3X}$, $\delta_{G4X}$, $\delta_{G5X}$ 
dominates over $\delta_{PX}$ during inflation, the relation 
between $r$ and $n_t$ is 
\begin{eqnarray}
r &=& -8.709\,n_t\,,\qquad (G_3~{\rm dominant})\,,\label{consis1}\\
r &=& -8.018\,n_t\,,\qquad (G_4~{\rm dominant})\,,\\
r &=& -8.095\,n_t\,,\qquad (G_5~{\rm dominant})\,.\label{consis3}
\end{eqnarray}
For the Galileon model in which only one of the $G_i$ terms 
($i=3,4,5$) is present, Eqs.~(\ref{scalarpower}), (\ref{nR}), and 
(\ref{ratio}) can be written in the forms \cite{Ohashi}
\begin{equation}
{\cal P}_{\cal R}=\frac{V^3}{12\pi^2 M_{\rm pl}^6 V_{,\phi}^2}f_i ({\cal A})\,,
\qquad
n_s-1=-6\epsilon_{V} g_{\epsilon i}({\cal A})+2\eta_V g_{\eta i}({\cal A})\,,
\qquad
r= 16\epsilon_V h_i ({\cal A})\,,
\label{Galiobser}
\end{equation}
where the functions $f_i$, $g_{\epsilon i}$, $g_{\eta i}$, and $h_i$ are
\begin{eqnarray}
& &f_3({\cal A})=\frac{(1+{\cal A})^2 (1+2{\cal A})^{1/2}}{(1+4{\cal A}/3)^{3/2}}\,,\quad
f_4({\cal A})=\frac{(1+{\cal A})^2 (1+3{\cal A})^{1/2}}{(1+13{\cal A}/9)^{3/2}}\,,\quad
f_5({\cal A})=\frac{(1+{\cal A})^2 (1+4{\cal A})^{1/2}}{(1+8{\cal A}/5)^{3/2}}\,,\\
& & g_{\epsilon 3}({\cal A})=g_{\epsilon 4}({\cal A})=g_{\epsilon 5}({\cal A})=
\frac{1}{1+{\cal A}}\,,\\
& & g_{\eta 3}({\cal A})=\frac{6+23{\cal A}+24{\cal A}^2}{2(1+2{\cal A})^2(3+4{\cal A})}\,,
\quad
g_{\eta 4}({\cal A})=\frac{9+46{\cal A}+78{\cal A}^2}{(1+3{\cal A})^2(9+13{\cal A})}\,,
\quad
g_{\eta 5}({\cal A})=\frac{5+31{\cal A}+80{\cal A}^2}{(1+4{\cal A})^2(5+8{\cal A})}\,,\\
& & h_{3}({\cal A})=\frac{(1+4{\cal A}/3)^{3/2}}{(1+{\cal A})^2(1+2{\cal A})^{1/2}}\,,\quad
h_{4}({\cal A})=\frac{(1+13{\cal A}/9)^{3/2}}{(1+{\cal A})^2(1+3{\cal A})^{1/2}}\,,\quad
h_{5}({\cal A})=\frac{(1+8{\cal A}/5)^{3/2}}{(1+{\cal A})^2(1+4{\cal A})^{1/2}}\,.
\end{eqnarray}

The amplitude ${\cal P}_{\cal R}$ is constrained to be 
${\cal P}_{\cal R} \simeq 2.2 \times 10^{-9}$ at $k_0=0.002$~Mpc$^{-1}$
from the Planck data \cite{Adeinf}. 
For the power-law potential (\ref{powerlaw}), 
this provides a relation between $\lambda$ and $M$
for a given value of $n$.

\subsection{Field-derivative couplings to the Einstein tensor}

The model of the field-derivative couplings to the Einstein 
tensor is given by the action \cite{Amendola93,Germani}
\be
{\cal S} = \int d^{4}x \sqrt{-g}\
\left[\frac{M_{\rm pl}^{2}}{2} R+X-V(\phi)
+\frac{1}{2M^2} G^{\mu \nu} \partial_{\mu} \phi
\partial_{\nu} \phi \right]\,,
\label{fieldde}
\ee
where $M$ is a constant having a dimension of mass.
In the Horndeski's action, this corresponds to the choice 
$G_5(\phi)=-\phi/(2M^2)$ 
(with integration by parts) \cite{KMY,Tsuji2012}.

{}From Eqs.~(\ref{epsi}) and (\ref{E1d}), the same equations as 
Eqs.~(\ref{epGa}),  (\ref{Hubble1}), and (\ref{field1}) 
hold with the replacement
\be
{\cal A} \equiv -\frac{6 \delta_{G5\phi}}{\delta_{PX}}
=\frac{3H^2}{M^2} \simeq \frac{V}{M^2 M_{\rm pl}^2}\,.
\ee
There is also the relation $\epsilon=\epsilon_V/(1+{\cal A})$.
The field value $\phi_f$ at the end of inflation and the number of 
e-foldings $N$ are known from Eqs.~(\ref{pend}) and 
(\ref{Nga}), respectively.

Since $q_s=\delta_{PX}-6 \delta_{G5\phi}$, the condition 
$q_s>0$ is automatically satisfied for $G_5=-\phi/(2M^2)$.
We also have $c_s^2=1$ at leading order of slow-roll.
Since $\epsilon_s=\epsilon=\delta_{PX}-6 \delta_{G5\phi}$, 
the consistency relation is \cite{GW}
\be
r=-8n_t\,,
\label{confi}
\ee
which is the same as that of standard slow-roll inflation.
Equations~(\ref{scalarpower}), (\ref{nR}), and (\ref{ratio}) can 
be expressed as \cite{Tsuji2012}
\begin{equation}
{\cal P}_{\cal R}=\frac{V^3}{12\pi^2 M_{\rm pl}^6 V_{,\phi}^2}
(1+{\cal A})\,,
\qquad
n_s-1=-6\epsilon_{V} \frac{1+4{\cal A}/3}{(1+{\cal A})^2}
+2\eta_V \frac{1}{1+{\cal A}}\,,
\qquad
r= \frac{16\epsilon_V}{1+{\cal A}}\,.
\label{fieldns}
\end{equation}
For the power-law potential (\ref{powerlaw}), 
$n_s$ and $r$ reduce to 
\begin{equation}
n_s=1-\frac{n^2[n(n+2)+2(n+1)\alpha x^n]}
{x^2 (n+\alpha x^n)^2}\,,\qquad
r=\frac{8n^3}{x^2 (n+\alpha x^n)}\,,
\end{equation}
where $\alpha=\lambda M_{\rm pl}^{n-2}/M^2$ and 
$x=\phi/M_{\rm pl}$. 
The number of e-foldings is given by 
\begin{equation}
N=\frac{x^2}{2n} \left[ 1+\frac{2\alpha}{n(n+2)} x^n \right]
-\frac{x_f^2}{2n} \left[ 1+\frac{2\alpha}{n(n+2)} x_f^n \right]\,,
\end{equation}
where $x_f$ is known by solving 
$2x_f^2 (1+\alpha x_f^n/n)=n^2$.

\subsection{k-inflation}

We consider the k-inflationary scenario in which the non-linear 
terms of $X$ are present in the Lagrangian \cite{kinf}. 
In the following, we focus on the models in which 
the scalar propagation speed $c_s$ is constant.
In fact, the constant values of $c_s$ appear in the context 
of power-law k-inflation ($a \propto t^p$ with constant $p$). 
Generally, such a power-law expansion can be realized 
for the Lagrangian \cite{Ohashi2011}
\begin{equation}
P(\phi,X)=X\,g(Y)\,,\qquad Y \equiv X e^{\lambda \phi}\,,
\label{scalinglag}
\end{equation}
where $g$ is an arbitrary function of $Y$, and $\lambda$
is a constant.
Originally, the Lagrangian (\ref{scalinglag}) was derived
for the existence of scaling solutions in the presence of
a barotropic fluid \cite{Piazza,Sami}.
Under the condition $\lambda^2<2P_{,X}$, there exists 
a power-law inflationary solution \cite{Tsuji06}. 
In the presence of multiple scalar fields with the Lagrangian 
$P=\sum_{i=1}^n X_i\,g(X_i e^{\lambda_i \phi_i})$, assisted 
inflation can be realized with the effective 
slope $\lambda=(\sum_{i=1}^n 1/\lambda_i^2)^{-1/2}$.
The simplest example is a canonical field 
with an exponential potential 
($g(Y)=1-c/Y$, i.e., $P=X-c e^{-\lambda \phi}$) \cite{Liddle}.
Thus, the Lagrangian (\ref{scalinglag}) not only realizes 
power-law inflation with constant $c_s$, but also 
it corresponds to the effective single-field Lagrangian of 
assisted inflation.

For the Lagrangian (\ref{scalinglag}), the background 
equations (\ref{E1d}) and (\ref{E1}) read
\begin{equation}
3M_{\rm pl}^2 H^2=X(g+2g_1)\,,\qquad
\epsilon=\frac{3(g+g_1)}{g+2g_1}\,,
\end{equation}
where $g_n \equiv Y^n (d^n g/dY^n)$. 
If we define the field equation of state as $w_{\phi}=P/(2XP_{,X}-P)$, 
the power-law inflationary solution mentioned above 
corresponds to $w_{\phi}=-1+\lambda^2/(3P_{,X})$ \cite{Tsuji06,Ohashi2011}.
Then, this solution satisfies the relation 
$\lambda=6(g+g_1)^2/(g+2g_1)$, along which 
$Y$ and $\epsilon$ are constants. 
{}From Eqs.~(\ref{nR}), (\ref{ratio}), and (\ref{nt}), we have 
\begin{eqnarray}
n_s-1 &=& -2\epsilon=n_t\,,
\label{nsk} \\
r &=& 16 c_s \epsilon=-8c_s n_t \,, 
\label{rk}
\end{eqnarray}
where the scalar propagation speed squared is
\begin{eqnarray}
c_s^2=\frac{g+g_1}{g+5g_1+2g_2}\,,
\end{eqnarray}
which is constant along the power-law inflationary 
solution. The leading-order non-linear estimator
(\ref{fnleq}) of the equilateral triangle ($k_1=k_2=k_3$) 
reads
\be
f_{\rm NL}^{\rm eq} = \frac{85}{324}
\left( 1-\frac{1}{c_s^2} \right)-\frac{10}{243}
\frac{6g_1+9g_2+2g_3}{g+5g_1+2g_2}\,.
\label{eqk}
\ee

For a given model (i.e., for a given form of $g(Y)$), 
the variable $Y$ is known in terms of $c_s$.
Then, the quantities (\ref{nsk}), (\ref{rk}),
and (\ref{eqk}) can be expressed with respect to $c_s$.
Note that, in order to exit from the regime of 
power-law k-inflation, the Lagrangian 
needs to be modified around the end of inflation.
We assume that for the scales relevant to the CMB anisotropies
the Lagrangian is well approximated by Eq.~(\ref{scalinglag}).
Our analysis does not exhaust all the possibilities of k-inflationary scenarios 
with time-varying $c_s$, but those general analyses are more 
complicated because the scalar spectral index involves 
the term $s$.

In the following, we consider two representative 
models with power-law k-inflation.

\subsubsection{Dilatonic ghost condensate}

The dilatonic ghost condensate model \cite{Piazza}, which 
is a generalization of the ghost condensate 
model \cite{Arkani}, is described by the Lagrangian
\begin{equation}
P=-X+c\,e^{\lambda \phi}X^2\,,
\label{dila}
\end{equation}
where $c$ is a constant. 
This model arises in low-energy effective string theory with $\alpha'$ 
corrections after the conformal transformation to the Einstein frame \cite{Piazza}.
By making a field redefinition \cite{Sami}, one can show that
this is equivalent to the model $P \propto \phi^{-2} (-X+X^2/M^4)$ 
first discussed in Ref.~\cite{kinf}.
The Lagrangian (\ref{dila}) corresponds to the function $g(Y)=-1+cY$ 
in Eq.~(\ref{scalinglag}).
Since $\epsilon=3(2cY-1)/(3cY-1)$, $c_s^2=(2cY-1)/(6cY-1)$, 
and $\mu/\Sigma=(1-c_s^2)/2$
in this case, it follows that 
\be
n_s-1=n_t=-\frac{24c_s^2}{1+3c_s^2}\,,\qquad
r=\frac{192c_s^3}{1+3c_s^2}\,,\qquad
f_{\rm NL}^{\rm eq}=-\frac{85}{324} \frac{1}{c_s^2}
+\frac{5}{81}c_s^2+\frac{65}{324}\,,
\label{nsrdbi}
\ee
which are written in terms of $c_s$ alone.

\subsubsection{DBI model}

The DBI model \cite{DBI} is given by the Lagrangian 
\begin{equation}
P=-f^{-1} (\phi) \sqrt{1-2f(\phi) X}
+f^{-1} (\phi)-V(\phi)\,,
\label{DBI}
\end{equation}
where $f(\phi)$ and $V(\phi)$ are functions of $\phi$.
If we choose the function 
$g(Y)=-(m^4/Y)\sqrt{1-2Y/m^4}-M^4/Y$ 
with two constant mass scales $m$ and $M$, 
we obtain the Lagrangian (\ref{DBI}) with  
\begin{equation}
f^{-1}(\phi)=m^4 e^{-\lambda \phi}\,,\qquad
V(\phi)=m^4(c_M+1)e^{-\lambda \phi}\,,\quad
{\rm where} \quad c_M \equiv M^4/m^4\,.
\label{DBIfun}
\end{equation}
In this case, we have 
\begin{equation}
\epsilon=\frac{3\tilde{Y}}{1+c_M \sqrt{1-2\tilde{Y}}}\,,
\qquad
c_s^2=1-2\tilde{Y}\,,\quad
{\rm where} \quad \tilde{Y} \equiv Y/m^4\,.
\label{DBIep}
\end{equation}
Using the relation $\mu/\Sigma=(1-c_s^2)/(2c_s^2)$,  
we obtain
\be
n_s-1=n_t=-\frac{3(1-c_s^2)}{1+c_M c_s}\,,\qquad
r=\frac{24c_s (1-c_s^2)}{1+c_M c_s}=8c_s(1-n_s)\,,\qquad
f_{\rm NL}^{\rm eq}=-\frac{35}{108} 
\left( \frac{1}{c_s^2}-1 \right)\,.
\label{dbifnl}
\ee

The ultra-relativistic regime ($c_s \ll 1$) corresponds to the case where
$\tilde{Y}$ is close to $1/2$. If $c_M$ is much larger than 1, 
the condition $c_M c_s \gg 1$ can be 
satisfied even for $c_s \ll 1$.
In this case, both $n_s-1$ and $r$ are much smaller than 1.
When $c_s \ll 1$, $|f_{\rm NL}^{\rm eq}|$ can be much larger than 1.

The functions (\ref{DBIfun}) are different from those 
in the original DBI model \cite{DBI}. 
The original version of the DBI model is at odds with 
observations even with the WMAP data \cite{DBIobs}.
We would like to study whether the DBI model with power-law
inflation can be consistent with the latest Planck data. 
Although our analysis based on the Lagrangian (\ref{scalinglag}) is 
phenomenological for the DBI model, it has an 
advantage to reduce the number of free parameters
appearing in inflationary observables.
We note that there exists another type of the DBI model 
characterized by the potential $V(\phi)=V_0-\beta H^2 \phi^2/2$, 
where $\beta$ is a constant with the theoretical allowed 
range $0.1<\beta<10^9$ \cite{IRDBI}.
In this case, the Planck group put a tight bound
$\beta \leq 0.7$ (95\,\%\,CL) by taking into account 
the constraint from the non-Gaussianities \cite{Adeinf}.

%%%%%%%%%%%%%%%%%%%%%%%%%%%%%%
\section{Joint observational constraints on 
single-field inflationary models}
\label{constsec}
%%%%%%%%%%%%%%%%%%%%%%%%%%%%%%

In this section, we put observational constraints on each 
inflationary model discussed in the previous section.
In doing so, we expand the power spectra ${\cal P}_{\cal R}$
and ${\cal P}_{h}$ around the pivot wave number $k_0$, as 
\ba
\ln {\cal P}_{\cal R}(k) &=& \ln {\cal P}_{\cal R}(k_0)
+\left[ n_{s}(k_0)-1 \right]y+
\frac{\alpha_{s}(k_0)}{2}y^2+{\cal O}(y^3)\,,
\label{Pexp1} \\
\ln {\cal P}_h (k) &=& \ln {\cal P}_h (k_0)
+n_t(k_0)y+\frac{\alpha_{t}(k_0)}{2}y^2
+{\cal O}(y^3)\,,
\label{Pexp2}
\ea
where $y=\ln (k/k_0)$. 
For the validity of the Taylor expansion, we require 
the following conditions
\be
\left| \alpha_s (k_0) \right|< 2\left| [n_s(k_0)-1]/y \right|\,,
\qquad
\left| \alpha_t (k_0) \right|< 2\left| n_t(k_0)/y \right|\,.
\label{criterion}
\ee
Under the slow-variation approximation, both $|\alpha_s (k_0)|$
and $|\alpha_t (k_0)|$ are of the order of $\epsilon^2$, 
whereas $|n_s(k_0)-1|$ and $|n_t (k_0)|$ are ${\cal O}(\epsilon)$.
The value $y$ depends on the choice of $k_0$.
For the scales relevant to the CMB anisotropies 
($2 \le l \lesssim 2500$), $y$ is smaller than 7. 
Then, the convergence criteria (\ref{criterion}) are well satisfied.

Following the Planck paper \cite{Adeinf}, we set both 
$\alpha_s(k_0)$ and $\alpha_t(k_0)$ to be 0
in the CMB likelihood analysis.
This is valid for the slow-variation inflationary models.
We confirmed that even with the prior $\alpha_{s,t}<10^{-3}$ the 
likelihood results are very similar to those in the case $\alpha_{s,t}=0$.
Setting the runnings to be 0, we are left with four parameters ${\cal P}_{\cal R}(k_0)$, 
$n_{s}(k_0)$, $r(k_0)$, and $n_t (k_0)$. 
If we specify the models, there are some relations 
between $n_{s}(k_0)$, $r(k_0)$, and $n_t (k_0)$.
This allows us to reduce the number of free parameters.

We take the pivot wave number to be 
\be
k_0=0.05~{\rm Mpc}^{-1}\,,
\label{k0}
\ee
which corresponds to a smaller scale relative to the value 
$k_0=0.002$~${\rm Mpc}^{-1}$ chosen by the Planck team.
Around the pivot scale (\ref{k0}) the CMB spectrum is
hardly affected by the uncertainty of cosmic variance. 
Note that we also performed the CosmoMC analysis for 
$k_0=0.002$~${\rm Mpc}^{-1}$
and reproduced the results presented in Ref.~\cite{Adeinf}.
We confirmed that the likelihood results are insensitive to 
the choice of different values of $k_0$.
We carry out the joint data analysis of
Planck \cite{Adecosmo}, WP \cite{WMAP9}, 
BAO \cite{BAO1,BAO2,BAO3}, 
and high-$\ell$ \cite{Das,Rei}.
In comparison, we also show the likelihood contours
constrained by the Planck+WP+BAO data.

In the whole analysis, the flat $\Lambda$CDM 
model is assumed with $N_{\rm eff}=3.046$ relativistic degrees 
of freedom \cite{Mangano}.
We employ the big bang nucleosynthesis consistency relation 
in that the helium fraction $Y_p$ is expressed in terms of 
$N_{\rm eff}$ and the baryon fraction $\Omega_b h^2$ \cite{Ichikawa}.
We also assume that reionization occurs instantly
at a redshift $z_{\rm re}$.

In the following, we proceed to observational constraints on 
each inflationary model.
For the scales relevant to the CMB anisotropies,
we fix the number of e-foldings to be $N=60$.
In principle, we can choose smaller values of $N$
like 50, but if some model is under an observational
pressure for $N=60$, it is typically more difficult to 
be compatible with observational constraints for $N=50$.
However, there are some exceptions, so we 
discuss such cases separately.

\subsection{Potential-driven slow-roll inflation}

For the potential-driven slow-roll inflation with the Lagrangian 
(\ref{standinf}), the consistency relation between $r(k_0)$ 
and $n_t(k_0)$ is given by $r(k_0)=-8n_t (k_0)$.
With the CosmoMC code, we perform the likelihood analysis 
by varying the three inflationary parameters ${\cal P}_{\cal R}(k_0)$, 
$n_{s}(k_0)$ and $r(k_0)$ together with other cosmological parameters. 

\begin{figure}
\includegraphics[height=3.9in,width=3.8in]{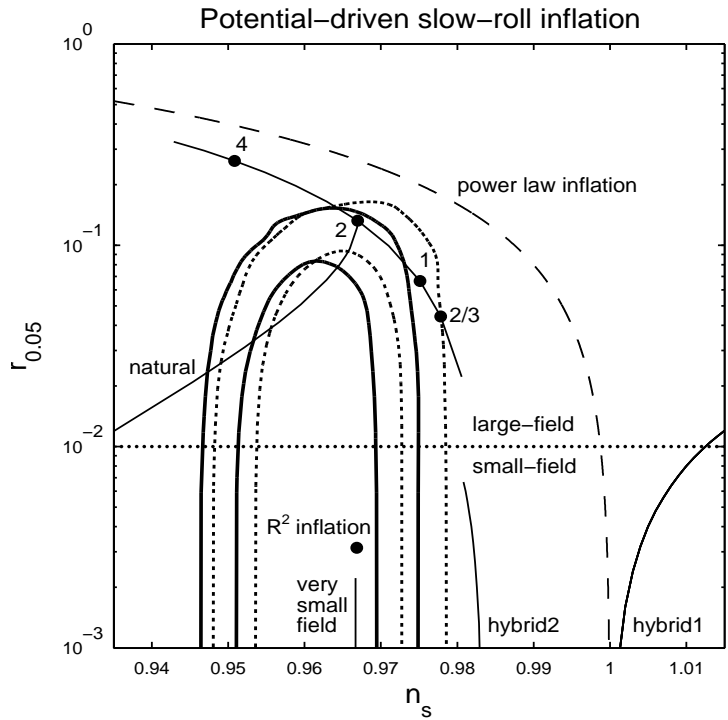}
\caption{\label{fig1}
2-dimensional observational constraints on potential-driven slow-roll inflation 
in the $(n_s,r)$ plane with the number of e-foldings $N=60$ and 
the pivot wave number $k_0=0.05$~Mpc$^{-1}$.
The bold solid curves represent the 68\,\%\,CL (inside) and 
95\,\%\,CL (outside) boundaries derived by 
the joint data analysis of Planck+WP+BAO+high-$\ell$, 
whereas the dotted curves correspond to the 68\,\% and 
95\,\% contours constrained by Planck+WP+BAO. 
In both cases the consistency relation 
$r(k_0)=-8n_t(k_0)$ is used.
We show the theoretical predictions for the 
models: (i) chaotic inflation with the potential 
$V(\phi)=\lambda \phi^n/n$ for general $n$ (thin solid curve) and 
for $n=4,2,1,2/3$ (denoted as black circles), (ii) natural inflation with 
the potential $V(\phi)=\Lambda^4 [1+\cos (\phi/f)]$ for general $f$, 
(iii) hybrid inflation with the potentials 
$V(\phi)=\Lambda^4+m^2 \phi^2/2$ (``hybrid1'') and 
$V(\phi)=\Lambda^4 [1+c \ln (\phi/\mu)]$ (``hybrid2''), 
(iv) very small-field inflation with the potential 
$V(\phi)=\Lambda^4 (1-e^{-\phi/M})$ in the regime 
$M<M_{\rm pl}$, and 
(v) power-law inflation with the exponential potential 
$V(\phi)=V_0 e^{-\gamma \phi/M_{\rm pl}}$.
The dotted line ($r=10^{-2}$) corresponds to the boundary
between ``large-field'' and ``small-field'' models.
For comparison, we also show the theoretical prediction
of the Starobinsky's model $f(R)=R+R^2/(6M^2)$ 
(denoted as ``$R^2$ inflation'').}
\end{figure}

The thick dotted curves in Fig.~\ref{fig1} correspond 
to the 68\,\%\, and 95\,\%\, CL boundaries
in the $(n_s, r)$ plane
constrained by the joint data analysis of Planck, WP, and BAO.
These bounds are similar to those derived by the Planck mission, 
in spite of the different choice of $k_0$
(see Fig.~1 in Ref.~\cite{Adeinf}).
The thick solid curves represent the 68\,\%\, and 95\,\%\, CL
borders derived by the joint data analysis of the 
Planck, WP, BAO, and high-$\ell$ data. 
Adding the high-$\ell$ data leads to the shift of 
$n_s$ toward smaller values and the slight decrease of $r$.
While the Planck group showed the bounds obtained from the 
Planck, WP, and high-$\ell$ data in Fig.~1 of Ref.~\cite{Adeinf}, 
we also included the BAO data.
The latter gives tighter bounds on $n_s$ than those constrained
by the former. 
In the following, we place observational constraints on 
each inflaton potential.

%%%%%%%%%%%%%%%%%%%%%%%%%%%%%%%%%%%
%%%%%%%%%%%%%%%%%%%%%%%%%%%%%%%%%

%
\subsubsection{Chaotic inflation}

For the power-law potential (\ref{powerlaw}), $n_s$ and $r$ are 
given by Eq.~(\ref{nsrstan}). 
The quartic potential ($n=4$) gives the values $n_s=0.951$ 
and $r=0.262$ for $N=60$, which is outside 
the 95\,\%\,CL contour\footnote{In the context of warm inflation
the quartic potential can be compatible with the Planck data \cite{Arjun}. }.
For the quadratic potential ($n = 2$) with $N = 60$, 
we have $n_s = 0.967$ and $r =0.132$.
This is close to the 95\,\%\,CL boundary constrained
by the Planck+WP+BAO+high-$\ell$ data.

The axion monodromy scenario \cite{Mca} gives rise to the linear potential 
with $n=1$. For $N=60$, this potential is within the 95\,\%\,CL region 
constrained by the Planck+WP+BAO data, but it is outside the 95\,\%\,CL
boundary by adding the high-$\ell$ data. 
This latter bound is tighter than that derived by the Planck team
based on the Planck, WP, and high-$\ell$ data. 
For $N=50$, however, we have $n_s=0.970$ and $r=7.96 \times 10^{-2}$, 
so that the linear potential enters the joint 95\,\%\,CL region
constrained by Planck+WP+BAO+high-$\ell$.
There is another power-law potential with $n=2/3$ appearing 
in axion monodromy \cite{West}. 
For $N=60$, this potential is outside the joint 95\,\%\,CL boundary
constrained by Planck+WP+BAO+high-$\ell$.
For $N=50$, we have $n_s=0.973$ and $r=5.32 \times 10^{-2}$, 
in which case the model marginally lies within the 
95\,\%\,CL contour. 
We note that there are some related models in which 
$r$ can be smaller than that discussed above \cite{Hagiwara,mflation}.

The exponential potential $V(\phi)=V_0 e^{-\gamma \phi/M_{\rm pl}}$, which 
gives rise to the power-law inflation $a \propto t^{2/\gamma^2}$, 
is characterized by the line (\ref{exp}) in the ($n_s, r$) plane. 
{}From Fig.~\ref{fig1}, we find that this model 
(denoted as a dashed curve) is excluded at more than 95\,\%\,CL.

\subsubsection{Natural inflation}

Natural inflation is characterized by the potential 
\be
V(\phi)=\Lambda^4 \left[ 1+\cos(\phi/f) \right]\,,
\label{naturalpo}
\ee
where $\Lambda$ and $f$ are constants having 
a dimension of mass.
If inflation occurs in the region around $\phi=0$, 
the expansion of Eq.~(\ref{naturalpo}) gives rise 
to the hill-top potential of the form $V(\phi)=2\Lambda^4 [1-\phi^2/(4f^2)+\cdots]$.
For the potential (\ref{naturalpo}), the number of e-foldings is related to 
the field $\phi$, as $N=(2f^2/M_{\rm pl}^2)\ln [\sin (\phi_f/(2f))/\sin (\phi/(2f))]$, 
where $\phi_f$ is known by solving $\tan^2 [\phi_f/(2f)]=2(f/M_{\rm pl})^2$.
The inflationary observables are
\be
n_s=1-\frac{M_{\rm pl}^2}{f^2} \frac{3-\cos(\phi/f)}
{1+\cos(\phi/f)}\,,\qquad r=\frac{8M_{\rm pl}^2}{f^2}
\frac{\sin^2(\phi/f)}{[1+\cos(\phi/f)]^2}\,.
\label{naobser}
\ee
For a given value of $f$, we can numerically identify the value of 
$\phi$ at  $N=60$. Then, $n_s$ and $r$ are evaluated from 
Eq.~(\ref{naobser}). 
In the limit that $f \to \infty$, inflation occurs
in the regime where $\phi$ is close to the potential minimum
($\phi=\pi f$). In this limit, $n_s$ and $r$ approach the values of 
chaotic inflation with $n=2$, i.e.,
$n_s=1-4/(2N+1)$ and $r=16/(2N+1)$.
For smaller $f$, both $n_s$ and $r$ tend to decrease.

In Fig.~\ref{fig1}, we show the theoretical values of $n_s$ and 
$r$ for different values of $f$.
There are intermediate values of $f$ with which the model
is within the 68\,\%\,CL region.
{}From the joint data analysis of Planck+WP+BAO+high-$\ell$,  
we obtain the following bounds
\ba
5.1M_{\rm pl}<&f&<7.9M_{\rm pl} \qquad  (68\,\%\,{\rm CL})\,,
\label{naturalcon0} \\
&f&>4.6 M_{\rm pl} \qquad(95\,\%\,{\rm CL})\,.
\label{naturalcon}
\ea
The bound (\ref{naturalcon}) is tighter than $f>3.5M_{\rm pl}$ 
derived in Ref.~\cite{Savage} with the WMAP data.

\subsubsection{Hybrid inflation}

Hybrid inflation is characterized by the potential 
\be
V(\phi)=\Lambda^4 +U(\phi)\,,
\label{hybridpo}
\ee
where $\Lambda$ is a constant, and $U(\phi)$
depends on $\phi$. 
Inflation ends due to the presence of a symmetry breaking 
field $\chi$. As long as $\chi$ is close to 0, 
the potential can be approximated by Eq.~(\ref{hybridpo})
during inflation.
The original hybrid model \cite{hybrid} corresponds to 
$U(\phi)=m^2 \phi^2/2$, in which case the curvature 
of the potential is positive ($V_{,\phi \phi}=m^2>0$).
There is a supersymmetric GUT model with 
$U(\phi)=c\,\Lambda^4 \ln (\phi/\mu)$ \cite{Dvali} (where 
$c$, $\Lambda$, $\mu$ are positive constants), 
in which case $V_{,\phi \phi}=-c\,\Lambda^4/\phi^2<0$.

We assume that the ratio $r_U \equiv U(\phi)/\Lambda^4$
is much smaller than 1.
For the potential $V(\phi)=\Lambda^4+m^2 \phi^2/2$, 
$n_s$ and $r$ are
\be
n_s \simeq 1+\frac{2m^2 M_{\rm pl}^2}{\Lambda^4}=1+2\eta_V\,,\qquad
r \simeq \frac{8m^4 M_{\rm pl}^2}{\Lambda^8}\phi^2 \simeq 8(n_s-1) r_{U}\,,
\ee
in which case $n_s>1$.
Under the condition $r_U<0.1$, the tensor-to-scalar ratio is 
constrained to be $r<0.8(n_s-1)$.
In Fig.~\ref{fig1}, the border $r=0.8 (n_s-1)$ corresponds 
to the solid curve in the regime $n_s>1$.
Clearly the hybrid model with $U(\phi)=m^2 \phi^2/2 \ll \Lambda^4$
is disfavored from the data.

For the potential $V(\phi)=\Lambda^4+c\,\Lambda^4 \ln (\phi/\mu)$ 
with $c \ll 1$, the number of e-foldings can be estimated as 
$N \simeq (\phi^2-\phi_c^2)/(2M_{\rm pl}^2 c)$, where 
$\phi_c$ is the field value at the bifurcation point.
Using the approximation $\phi^2 \gg \phi_c^2$, 
it follows that 
\be
n_s \simeq 1-\frac{2+3c}{2N}\,,\qquad
r \simeq \frac{4c}{N} \simeq \frac{8c}{2+3c} (1-n_s)\,,
\label{gutns}
\ee
in which case $n_s<1$.
For $0<c<0.1$ and $N=60$, the observables (\ref{gutns}) are 
in the ranges $0.9808<n_s<0.9833$ and 
$0<r<6.67 \times 10^{-3}$.
As we see in Fig.~\ref{fig1}, the theoretical curve lies outside
the 95\,\%\,CL region. 
Even for $N=50$ (which gives smaller values of $n_s)$, 
the model is outside the 95\,\%\,CL boundary constrained
by the Planck, WP, BAO, and high-$\ell$ data.

\subsubsection{Very small-field inflation}

There are some models in which the variation of the field  
during inflation is much smaller than $M_{\rm pl}$ and 
hence $r \ll 0.01$ from Eq.~(\ref{Lyth}).
Let us consider the inflaton potential of the form
\be
V(\phi)=\Lambda^4 [1-f(\phi)]\,,
\ee
where $\Lambda$ is a constant and
$f(\phi)$ is a function of $\phi$.
The function $f(\phi)=e^{-\phi/M}$ appears in the context of 
D-brane inflation \cite{Tye}.
The K\"{a}hler-moduli inflation corresponds to the choice
$f(\phi)=c_1 \phi^{4/3}e^{-c_2\phi^{4/3}}$ 
($c_1>0$, $c_2>0$) \cite{Conlon} (see also Refs.~\cite{Cicoli} 
for related works),
In both models, the potential $V(\phi)$ is asymptotically flat
in the limit $\phi \to \infty$.
There are other potentials of the form $f(\phi)=(M/\phi)^n$ 
($n>0$) \cite{Tye,KKLT} or $f(\phi)=\lambda_1 (\phi-\phi_0)
+\lambda_3 (\phi-\phi_0)^3/3!$ \cite{Baumann}, but these models 
are generally plagued by the so-called $\eta$-problem for the natural 
parameters constrained by string theory.

Let us consider the case $f(\phi)=e^{-\phi/M}$.
Then, the number of e-foldings is estimated as 
$N \simeq (M/M_{\rm pl})^2 e^{\phi/M}$.
We also obtain 
\be
n_s \simeq 1-\frac{2}{N}\,,\qquad  r \simeq \frac{8}{N^2}
\left( \frac{M}{M_{\rm pl}} \right)^2\,.
\ee
For $N=60$ and $M<M_{\rm pl}$, it follows that 
$n_s \simeq 0.967$ and $r<2.2 \times 10^{-3}$.
As we see in Fig.~\ref{fig1}, this model is inside 
the 68\,\%\,CL boundary constrained
by the Planck, WP, BAO, and high-$\ell$ data.

In K\"{a}hler-moduli inflation with 
$f(\phi)=c_1 \phi^{4/3}e^{-c_2\phi^{4/3}}$,
we have that $0.960<n_s<0.967$ and $r<10^{-10}$
for $50<N<60$ \cite{Conlon}. 
This model is well inside 
the 68\,\%\,CL contour.

\subsection{Non-minimally coupled models}

We study the non-minimally coupled models given by 
the action (\ref{actionsta}) with $F(\phi)=1-\xi \phi^2/M_{\rm pl}^2$ 
and $\omega=1$. 
Since the same consistency relation as that in standard 
potential-driven inflation ($r=-8n_t$) holds, the observational 
constraints in the $(n_s,r)$ plane are the same as those 
in Fig.~\ref{fig1}.
For two potentials $V(\phi)=m^2 \phi^2/2$ and 
$V(\phi)=\lambda \phi^4/4$,  we numerically evaluate 
the observables (\ref{nRex2}) and (\ref{rex2}) for the values
of $x=\phi/M_{\rm pl}$ corresponding to $N=60$.
We focus on the negative non-minimal 
couplings\footnote{In our notation the conformal 
coupling corresponds to $\xi=1/6$.}
with which $r$ gets smaller relative 
to the case $\xi=0$.

%%%%%%%%%%%%%%%%%%%%%%%%%%%%%%%%%%%
\begin{figure}
\includegraphics[height=3.7in,width=3.7in]{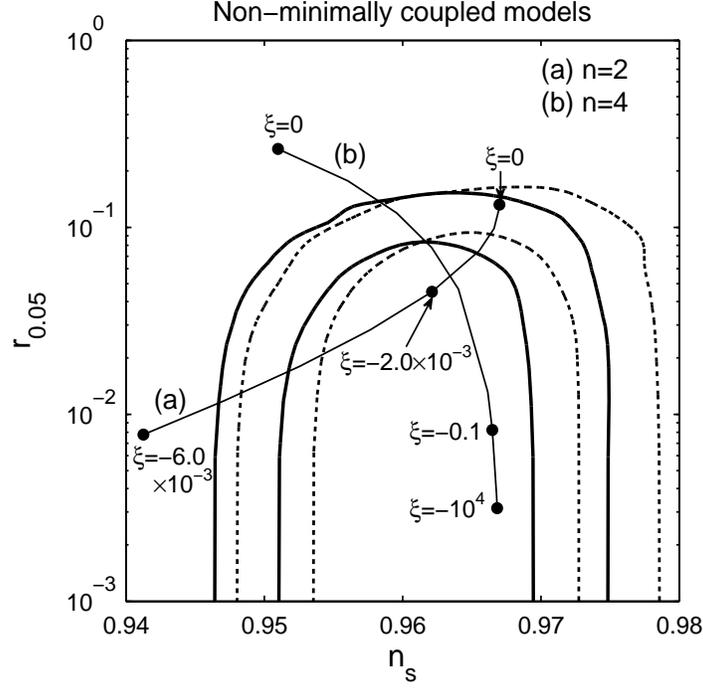}
\caption{\label{fig2}
2-dimensional observational constraints on non-minimally 
coupled models ($\xi R\phi^2/2$) 
with $N=60$. The 68\,\% and 95\,\%\,CL observational 
contours are the same as those given in Fig.~\ref{fig1}. 
The curves (a) and (b) show the theoretical 
predictions of the potentials $V(\phi)=m^2 \phi^2/2$ and 
$V(\phi)=\lambda \phi^4/4$, respectively, with negative 
values of $\xi$. For larger $|\xi|$, the scalar spectral 
index of the model (a) gets smaller. 
With the increase of $|\xi|$, the model (b) enters the region 
inside the 68\,\%\,CL boundary.
}
\end{figure}
%%%%%%%%%%%%%%%%%%%%%%%%%%%%%%%%%

For the potential $V(\phi)=m^2 \phi^2/2$, the scalar spectral 
index decreases for larger values of 
$|\xi|$ \cite{Gumjudpai,Lindenon,Reza}.
In Fig.~\ref{fig2}, we plot the theoretical values of $n_s$ 
and $r$ as a function of $\xi$.
{}From the joint data analysis of Planck+WP+BAO+high-$\ell$, 
we obtain the following bounds for $N=60$ : 
\ba
-4.2 \times 10^{-3}<&\xi&<-1.1 \times 10^{-3} \qquad  (68\,\%\,{\rm CL})\,,
\label{n=2con0}\\
-5.1 \times 10^{-3}<&\xi& \le 0 \qquad \qquad \qquad ~~\,(95\,\%\,{\rm CL})\,.
\label{n=2con}
\ea
The bound (\ref{n=2con}) is slightly tighter than
$\xi>-7.0 \times 10^{-3}$ (95\,\%\,CL) \cite{Reza} derived from 
the WMAP7 data with $N=55$.

For the potential $V(\phi)=\lambda \phi^4/4$, the negative non-minimal 
couplings lead to the increase of $n_s$ as well as the decrease 
of $r$ \cite{Komatsuper,Gumjudpai,Lindenon,Reza,Okada}.
In the limit that $|\xi| \to \infty$ with $N \gg 1$, the observables 
(\ref{nRex2}) and (\ref{rex2}) reduce to
\ba
n_s \simeq 1-\frac{2}{N}\,,\qquad  r \simeq \frac{12}{N^2}\,.
\label{nonlimit}
\ea
For $N=60$, we have $n_s=0.967$ and $r=3.33 \times 10^{-3}$, 
in which case the model is inside the 68\,\%\,CL contour.
In this regime, the scalar power spectrum (\ref{Psam}) is approximately 
given by ${\cal P}_{\cal R} \simeq \lambda N^2/(72 \pi^2 \xi^2)$.
Since the best-fit value of the scalar amplitude at $k_0=0.05$~Mpc$^{-1}$
is ${\cal P}_{\cal R}=2.2 \times 10^{-9}$, it follows that 
$\lambda/\xi^2 \simeq 4.3 \times 10^{-10}$.
For the self-coupling $\lambda=0.1$, we have
$\xi \simeq -1.5 \times 10^{4}$.
In Fig.~\ref{fig2}, the theoretical values of $n_s$ and $r$
are plotted as a function of $\xi$. 
The bounds on $\xi$ derived from the joint data analysis 
of Planck+WP+BAO+high-$\ell$ (for $N=60$) are
\ba
& &\xi<-4.5 \times 10^{-3} \qquad  (68\,\%\,{\rm CL})\,,
\label{n=4con0}\\
& &\xi<-1.9 \times 10^{-3} \qquad  (95\,\%\,{\rm CL})\,,
\label{n=4con}
\ea
which are slightly tighter than those
derived in Refs.~\cite{Reza,Okada} with the WMAP7 data.
These results are consistent with those of the Planck 
group \cite{Adeinf}.

\subsection{Running kinetic couplings}
%

%%%%%%%%%%%%%%%%%%%%%%%%%%%%%%%%%
\begin{figure}
\includegraphics[height=3.7in,width=3.7in]{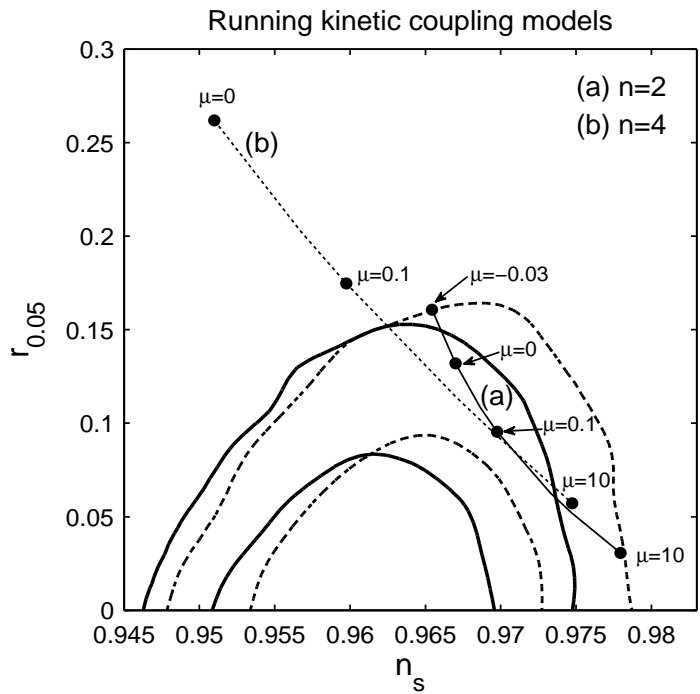}
\caption{\label{fig3}
2-dimensional observational constraints on the running kinetic 
coupling model ($\omega(\phi)=e^{\mu \phi/M_{\rm pl}}$) 
with $N=60$. 
The 68\,\% and 95\,\%\,CL observational 
contours are the same as those given in Fig.~\ref{fig1}. 
The curves (a) and (b) correspond to the cases of 
the potentials $V(\phi)=m^2 \phi^2/2$ and 
$V(\phi)=\lambda \phi^4/4$, respectively, 
with $\mu$ ranging $-0.03 \le \mu \le 10$ ($n=2$) 
and $0 \le \mu \le 10$ ($n=4$). 
For larger $\mu$, the scalar spectral index increases, 
while the tensor-to-scalar ratio gets smaller.
}
\end{figure}
%%%%%%%%%%%%%%%%%%%%%%%%%%%%%%%%%

We proceed to the running kinetic coupling model described 
by the action  (\ref{actionsta}) with $F(\phi)=1$ and 
$\omega(\phi)=e^{\mu \phi/M_{\rm pl}}$.
For the power-law potential (\ref{powerlaw}), the scalar 
spectral index and the tensor-to-scalar ratio 
are given by Eq.~(\ref{runningns}). 
For increasing $\mu$, $n_s$ gets larger, whereas
$r$ decreases.
If $\mu$ is much larger than 1, these observables 
can be estimated as \cite{Reza}
\be
n_s \simeq 1-\frac{1}{N}\,,\qquad
r \simeq \frac{8n}{N} \frac{1}{\mu x}\,,
\ee
where $N \simeq x e^{\mu x}/(n \mu)$.
In the limit $\mu \to \infty$,
the asymptotic values of $n_s$ and $r$ 
are $n_s \to 0.983$ and $r \to 0$ for $N=60$, respectively.
As we see in Fig.~\ref{fig3}, the running kinetic coupling 
model in this limit is outside the 95\,\%\,CL contour.

For the intermediate values of $\mu$, however, the models 
can be within the 95\,\%\,CL region even for the 
potential $V(\phi)=\lambda \phi^4/4$.
{}From the joint data analysis of Planck+WP+BAO+high-$\ell$, 
we find that $\mu$ is constrained to be 
\ba
-0.02<&\mu&<0.57\quad (95\,\%\,{\rm CL})\quad {\rm for}~n=2\,,
\label{runningn2}\\
0.18<&\mu&<5.0\quad~\, (95\,\%\,{\rm CL})\quad {\rm for}~n=4\,.
\label{runningn4}
\ea
These lower bounds of $\mu$ are slightly tighter than those 
derived in Ref.~\cite{Reza} with the WMAP7 data. 
The WMAP7 data do not put upper bounds of $\mu$
because even the model with $n_s=0.983$
and $r=0$ is allowed. 
With the Planck data, however, the tighter upper limits 
of $n_s$ constrain $\mu$ from above.

\subsection{Brans-Dicke theories}

We study inflation in Brans-Dicke theories described by the action 
(\ref{Jframe}) with the potential (\ref{poBD}). We recall that 
the potential (\ref{poBD}) recovers the Starobinsky's $f(R)$ model 
for $\omega_{\rm BD}=0$.
This approach is phenomenological, but it also has an advantage that the potential 
(\ref{Uchi}) in the Einstein frame (with $\gamma=1/\sqrt{3/2+\omega_{\rm BD}}$) 
can recover the quadratic potential $U(\chi) \simeq 3\gamma^2 M^2 \chi^2/4$ of 
chaotic inflation in the limit $\omega_{\rm BD} \gg 1$ (i.e., $\gamma \ll 1$).
In this limit, $n_s$ and $r$ approach the values (\ref{nsrstan})
of chaotic inflation with $n=2$.

%%%%%%%%%%%%%%%%%%%%%%%%%%%%%%%%%
\begin{figure}
\includegraphics[height=3.7in,width=3.7in]{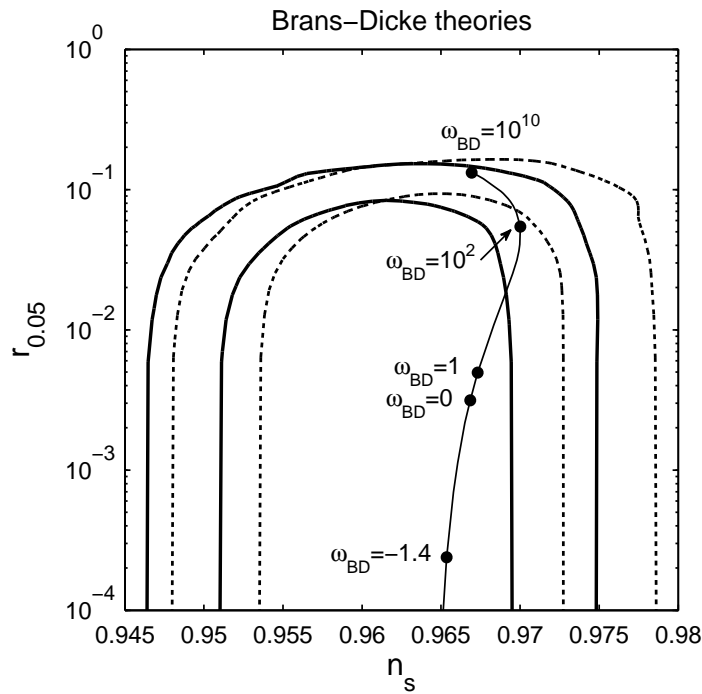}
\caption{\label{fig4}
2-dimensional observational constraints on inflation in 
Brans-Dicke theories 
in the presence of the potential (\ref{poBD}) with $N=60$. 
The 68\,\% and 95\,\%\,CL observational 
contours are the same as those given in Fig.~\ref{fig1}. 
The solid curve shows the theoretical predictions in 
the regime $-1.5<\omega_{\rm BD}<\infty$.
The Starobinsky's $f(R)$ model $f(R)=R+R^2/(6M^2)$ 
corresponds to the case $\omega_{\rm BD}=0$.
}
\end{figure}
%%%%%%%%%%%%%%%%%%%%%%%%%%%%%%%%%

For the theories with $|\omega_{\rm BD}| \lesssim {\cal O}(1)$, 
it follows that $F_f=1+\sqrt{2} \gamma={\cal O}(1)$ and 
$N \simeq F/(2 \gamma^2)$. Since $F \gg 1$ for $N \gg 1$, 
the observables (\ref{Bransns}) reduce to
\ba
n_s \simeq 1-\frac{2}{N}\,,\qquad
r \simeq \frac{4(3+2\omega_{\rm BD})}{N^2}\,.
\ea
The Starobinsky's model $f(R)=R+R^2/(6M^2)$ corresponds 
to $\omega_{\rm BD}=0$, in which case $n_s$ and $r$ are
the same as Eq.~(\ref{nonlimit}).
As we see in Fig.~\ref{fig4}, the tensor-to-scalar ratio 
decreases for smaller $\omega_{\rm BD}$ with the 
asymptotic value $r \to 0$ in the limit 
$\omega_{\rm BD} \to -3/2$.
{}From the joint data analysis of Planck+WP+BAO+high-$\ell$, 
the Brans-Dicke parameter 
is constrained to be 
\be
\omega_{\rm BD}<11.5 \qquad (68\,\%\,{\rm CL})\,.
\label{omecon}
\ee
\subsection{Potential-driven Galileon inflation}

We proceed to the potential-driven Galileon inflation described 
by the Lagrangian (\ref{Galilag}). 
We consider the effects of three covariant Galileon terms 
$G_3=c_3X/M^3$, $G_4=-c_4X^2/M^6$, and 
$G_5=3c_5X^2/M^9$ separately.  
We study the cases of two power-law potentials 
$V(\phi)=\lambda \phi^n/n$ with $n=2$ and $n=4$
in the regime $\phi>0$. Since $\dot{\phi}<0$ during inflation, 
we can choose the coefficients of the terms $G_i$ ($i=3,4,5$) to be 
$c_3=-1$, $c_4=-1$, and $c_5=-1$ without loss of 
generality (the signs of $c_i$
are fixed to avoid scalar ghosts).
For small $M$, there appears a regime in which 
the Galileon self-interaction dominates over the standard
kinetic term during inflation. 
In the limit that $M \to 0$, the scalar spectral 
index and the tensor-to-scalar ratio 
in Eq.~(\ref{Galiobser}) reduce to \cite{Kamada,Ohashi}
\ba
& &n_s=1-\frac{3(n+1)}{(n+3)N+n} \,, \qquad 
r=\frac{64 \sqrt{6}}{9}\frac{n}{(n+3)N+n} \qquad 
(G_3~{\rm dominant})\,,
\label{galiG3}\\
& &n_s=1-\frac{2(5n+4)}{4(n+2)N+3n}\,,\qquad
r=\frac{208 \sqrt{39}}{27} 
\frac{n}{4(n+2)N+3n} \qquad (G_4~{\rm dominant})\,,\\
& &n_s=1-\frac{7n+5}{(3n+5)N+2n}\,,\qquad
r=\frac{256 \sqrt{10}}{25} 
\frac{n}{(3n+5)N+2n} \qquad (G_5~{\rm dominant})\,.
\label{galiG5}
\ea

\begin{figure}
\includegraphics[height=3.7in,width=3.7in]{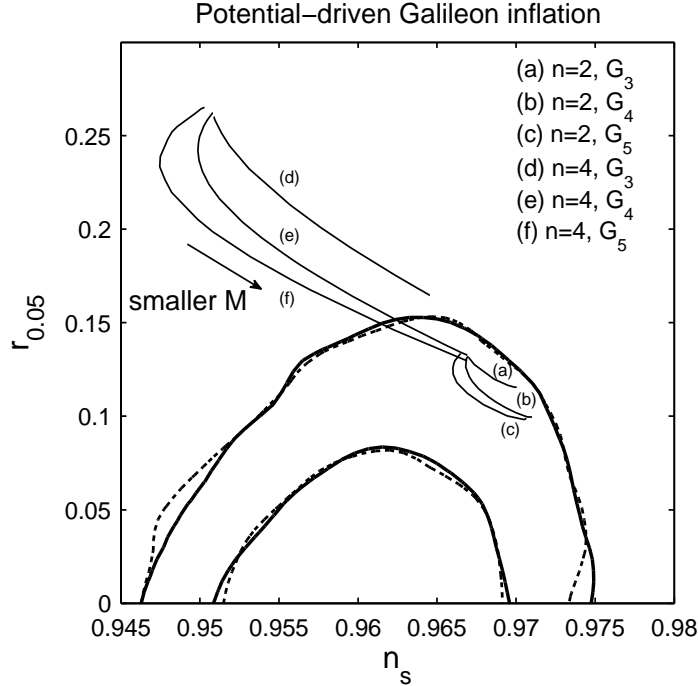}
\caption{\label{fig5}
2-dimensional observational constraints on potential-driven 
Galileon inflation with $N=60$ derived from the joint data 
analysis of Planck+WP+BAO+high-$\ell$. 
The bold solid curves correspond to the 
68\,\%\,CL (inside) and 95\,\%\,CL (outside)
contours derived by using the consistency relation $r=-8n_t$, 
whereas the dashed curves show the 
68\,\% and 95\,\%\,CL boundaries obtained from 
the consistency relation $r=-8.709n_t$.
Each thin solid curve shows the theoretical prediction of
the potentials $V(\phi)=m^2 \phi^2/2$ or 
$V(\phi)=\lambda \phi^4/4$ in the presence of 
the terms $G_3=-X/M^3$, or, $G_4=X^2/M^6$, 
or $G_5=-3X^2/M^9$. 
For smaller $M$, the tensor-to-scalar ratio decreases.}
\end{figure}

We recall that, in this regime, the consistency relations are
given by Eqs.~(\ref{consis1})-(\ref{consis3}) respectively, 
which are different from 
$r=-8n_t$ in standard slow-roll inflation.
In Fig.~\ref{fig5}, we plot the 68\,\% and 95\,\%\,CL observational 
contours derived by using the two consistency relations: 
$r=-8n_t$ and $r=-8.709n_t$.
Since they are similar to each other, we can safely 
use the observational bounds in the $(n_s, r)$ plane
constrained from the standard relation $r=-8n_t$.

As we see in Fig.~\ref{fig5}, $n_s$ increases for smaller $M$, 
whereas $r$ gets smaller.
For the potential $V(\phi)=m^2 \phi^2/2$ the theoretical curves 
lie inside the 95\,\%\,CL boundary, but still they are outside the 
68\,\%\,CL contour.
If the Galileon self-interaction dominates over the standard 
kinetic term even after inflation, there is no oscillatory regime 
of inflaton during reheating \cite{Ohashi}.
This is typically accompanied by the appearance of 
the negative propagation speed squared $c_s^2$.
This puts lower bounds on $M$, e.g., 
$M>4.2 \times 10^{-4}M_{\rm pl}$ for $G_3=-X/M^3$. 
Even for the values of $M$ around these lower bounds,  
$n_s$ and $r$ are close to the asymptotic values given 
in Eqs.~(\ref{galiG3})-(\ref{galiG5}).

For the potential $V(\phi)=\lambda \phi^4/4$ with the term $G_3=-X/M^3$, 
the asymptotic values in Eq.~(\ref{galiG3}) are 
$n_s=0.965$ and $r=0.164$ for $N=60$.
Because of the large tensor-to-scalar ratio, this model is 
outside the 95\,\%\,CL contour for arbitrary values of $M$. 
In the presence of the terms $G_4=X^2/M^6$ or
$G_5=-3X^2/M^9$, there are some values of $M$ 
with which the potential $V(\phi)=\lambda \phi^4/4$
enters the 95\,\%\,CL region. 
{}From the joint analysis of Planck+WP+BAO+high-$\ell$, 
we obtain the following bounds 
\ba
M &<&
7.2\times10^{-4} M_{\rm pl} \qquad (95\,\%\,{\rm CL})\qquad 
(G_4~{\rm dominant})\,,
\label{galicon1}\\
M&<&
6.6\times10^{-4} M_{\rm pl}  \qquad (95\,\%\,{\rm CL})\qquad 
(G_5~{\rm dominant})\,,
\label{galicon2}
\ea
which are tighter than those derived with the WMAP7 data: 
$M<1.1 \times 10^{-3} M_{\rm pl}$ ($G_4$ dominant) and 
$M<8.6\times10^{-4} M_{\rm pl}$ ($G_5$ dominant) \cite{Ohashi}.
In order to avoid the negative values of $c_s^2$, we require that 
$M>2.3 \times 10^{-4}M_{\rm pl}$ ($G_4$ dominant) and 
$M>2.9 \times 10^{-4}M_{\rm pl}$ ($G_5$ dominant) \cite{Ohashi}.
Hence there are still allowed parameter spaces 
compatible with the bounds (\ref{galicon1}) and (\ref{galicon2}).

%%%%%%%%%%%%%%%%%%%%%%%%%%%%%%%%%
%%%%%%%%%%%%%%%%%%%%%%%%%%%%%%%%%

%
\subsection{Field-derivative couplings to the Einstein tensor}
%

%%%%%%%%%%%%%%%%%%%%%%%%%%%%%%%%%
\begin{figure}
\includegraphics[height=3.7in,width=3.7in]{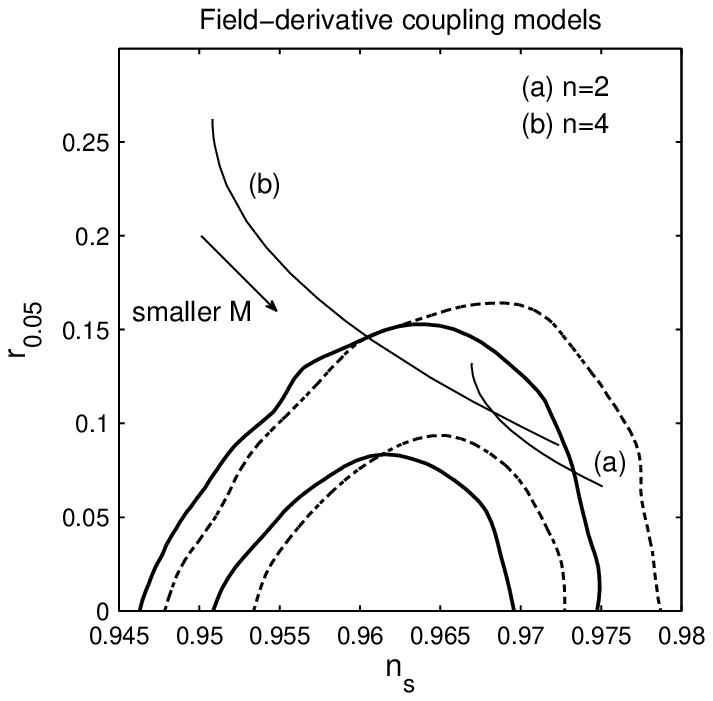}
\caption{\label{fig6}
2-dimensional observational constraints on field-derivative coupling models
with $N=60$. The 68\,\% and 95\,\%\,CL observational 
contours are the same as those given in Fig.~\ref{fig1}.   
The thin-solid curves correspond to the theoretical predictions of
the potentials $V(\phi)=m^2 \phi^2/2$ and  
$V(\phi)=\lambda \phi^4/4$, respectively, 
in the presence of the term 
$G^{\mu \nu} \partial_{\mu} \phi \partial_{\nu} \phi/(2M^2)$.
The tensor-to-scalar ratio gets smaller for decreasing $M$.
}
\end{figure}
%%%%%%%%%%%%%%%%%%%%%%%%%%%%%%%%%

The model with the field derivative couplings to the Einstein tensor
is given by the action (\ref{fieldde}).
In this case, the consistency relation (\ref{confi}) is the same
as that in standard slow-roll inflation.
In the following, we focus on the two potentials 
$V(\phi)=\lambda \phi^n/n$ with $n=2$ and $n=4$.
In the limit that $M \to 0$ (i.e., $\alpha=\lambda M_{\rm pl}^{n-2}/M^2 \to \infty$), 
the observables in Eq.~(\ref{fieldns}) reduce to \cite{Tsuji2012}
\be
n_s=1-\frac{4(n+1)}{2(n+2)N+n}\,,\qquad
r=\frac{16n}{2(n+2)N+n}\,.
\label{deriasy}
\ee
When $N=60$, we have $n_s=0.975$, $r=0.066$ for $n=2$
and $n_s=0.972$, $r=0.088$ for $n=4$.

In Fig.~\ref{fig6}, we plot the theoretical values of 
$n_s$ and $r$ as a function of $M$ for $n=2$ and 
$n=4$. Even in the presence of the field-derivative couplings,
both potentials are outside the 68\,\%\,CL region.
For the potential $V(\phi)=m^2 \phi^2/2$, the model with 
the asymptotic values (\ref{deriasy}) lies outside the 95\,\%\,CL
region constrained by the Planck, WP, BAO, and high-$\ell$ data.
This puts an upper bound on the parameter 
$\alpha$, as $\alpha<0.3$ (95\,\%\,CL).
On the other hand, from the joint analysis of Planck+WP+BAO, 
the theoretical curve is still inside the 
95\,\%\,CL contour.

For the potential $V(\phi)=\lambda \phi^4/4$, the model with the asymptotic 
values (\ref{deriasy}) is marginally inside the 95\,\%\,CL boundary
constrained from the Planck, WP, BAO, and high-$\ell$ data.
This potential is within the 95\,\%\,CL region for 
\be
\alpha=\lambda M_{\rm pl}^2/M^2> 9.0 \times 10^{-5}\,.
\label{alcon}
\ee
This bound is tighter than $\alpha>3.0 \times 10^{-5}$
derived by using the WMAP7 data \cite{Tsuji2012}.

\subsection{k-inflation}
\subsubsection{Dilatonic ghost condensate}
%

%%%%%%%%%%%%%%%%%%%%%%%%%%%%%%%%%
\begin{figure}
\includegraphics[height=3.4in,width=3.2in]{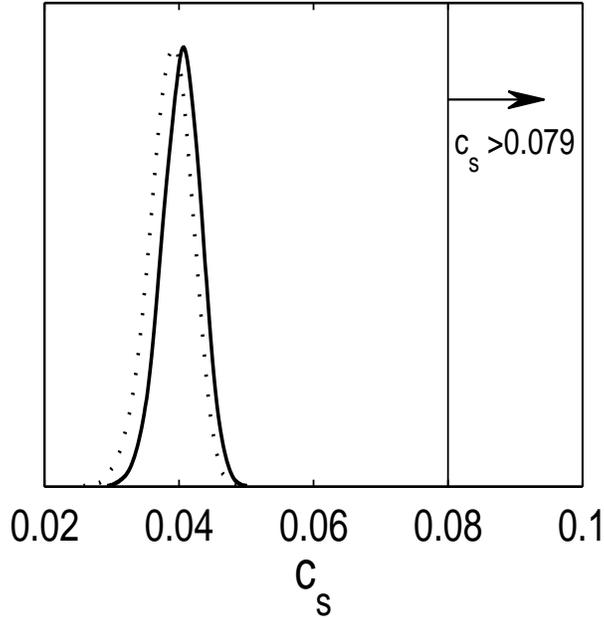}
\caption{\label{fig7}
1-dimensional marginalized probability distribution (solid curve)
of the scalar propagation speed $c_s$ constrained by 
the Planck, WP, BAO, and high-$\ell$ data 
in the dilatonic ghost condensate model. 
The dotted curve shows the probability distribution of 
$c_s$ derived from the Planck, WP, and BAO data.
The analysis of the non-Gaussianity puts the further bound 
$c_s>0.079$, by which the model is  
excluded at more than 95\,\%\,CL.}
\end{figure}
%%%%%%%%%%%%%%%%%%%%%%%%%%%%%%%%%

The dilatonic ghost condensate model is given by the Lagrangian 
(\ref{dila}). Using the formulas of $n_s$, $n_t$, and $r$ given 
in Eq.~(\ref{nsrdbi}), we carry out the likelihood analysis 
by varying $c_s$ in the range $0 \le c_s \le 1$ 
together with other cosmological parameters.
In Fig.~\ref{fig7}, we plot the 1-dimensional marginalized probability 
distributions of $c_s$ derived from the joint data analyses
of Planck+WP+BAO+high-$\ell$ (solid curve) and 
Planck+WP+BAO (dotted curve).
The Planck+WP+BAO+high-$\ell$ data give the following constraints 
\ba
& & 0.038 < c_s < 0.043 \qquad (68\,\%\,{\rm CL})\,,
\label{cscon2d}\\
& & 0.034 <c_s<0.046 \qquad (95\,\%\,{\rm CL})\,.
\label{cscon2}
\ea
We also obtain similar bounds from the joint analysis of Planck+WP+BAO.
Since the Harrison-Zeldovich spectrum ($n_s=1$, $r=0$) is disfavored from 
the data, the model with $c_s=0$ is outside the 95 \%\,CL boundary.

For $c_s \ll 1$, the non-linear estimator 
at the equilateral triangle given in Eq.~(\ref{nsrdbi}) satisfies 
$|f_{\rm NL}^{\rm eq}| \gg 1$. 
Hence there should be a lower bound of $c_s$.
In order to compare the model prediction with the observations 
of non-Gaussianities, we need to employ the leading-order bispectrum 
given by Eq.~(\ref{bispectrum}).
For the dilatonic ghost condensate model 
the bispectrum reads
\be
{\cal A}_{\cal R}^{\rm lead} =-\frac{13}{288} \left(\frac{1}{c_s^2}-1 \right)
\left[2+3\beta+c_s^2 (3\beta-2) \right]
S_7^{\rm equil}+\frac{1}{96} \left(\frac{1}{c_s^2}-1 \right)
(1+c_s^2) (13\beta-14)
S_7^{\rm ortho}\,.
\ee
In the limit $c_s^2 \ll 1$, it follows that 
${\cal A}_{\cal R} \simeq 
(-0.252/c_s^2)S_7^{\rm equil}+0.016S_7^{\rm ortho}$ 
and hence the equilateral shape dominates over 
the orthogonal one. 
The Planck team used the equilateral template 
introduced in Ref.~\cite{Paolo} for the 
above model (equivalent to ``power-law k-inflation'' of Ref.~\cite{Adenon})
and derived the following bound 
\be
c_s>0.079\qquad (95\,\%\,{\rm CL}).
\ee
This is not compatible with the constraint (\ref{cscon2}), 
so the dilatonic ghost condensate model is severely disfavored.
The same conclusion was reached in Ref.~\cite{Adenon}, 
but we derived more precise bounds (\ref{cscon2d})-(\ref{cscon2})
by performing the joint data analysis of Planck+WP+BAO+high-$\ell$.

\subsubsection{DBI model with power-law inflation}

The DBI model with power-law inflation is given by the 
Lagrangian (\ref{DBI}) with the functions (\ref{DBIfun}).
Using the formulas (\ref{dbifnl}), we vary two parameters 
$c_s$ and $n_s$ in the ranges $0< c_s \le 1$ and 
$0.90 \le n_s \le 1$.
When $c_s=0$, we have $\tilde{Y}=1/2$ and 
$\epsilon=3/2$ from Eq.~(\ref{DBIep}), in which case
inflation is not realized. 
Even for $c_s \ll 1$, as long as $c_s$ is not exactly 0, it is possible 
to find the large parameter $c_M$ satisfying the condition $c_M c_s \gg 1$.
Then, $|n_s-1|$ can be much smaller than 1 to match with the 
observational data.

%%%%%%%%%%%%%%%%%%%%%%%%%%%%%%%%%
\begin{figure}
\includegraphics[height=3.5in,width=3.5in]{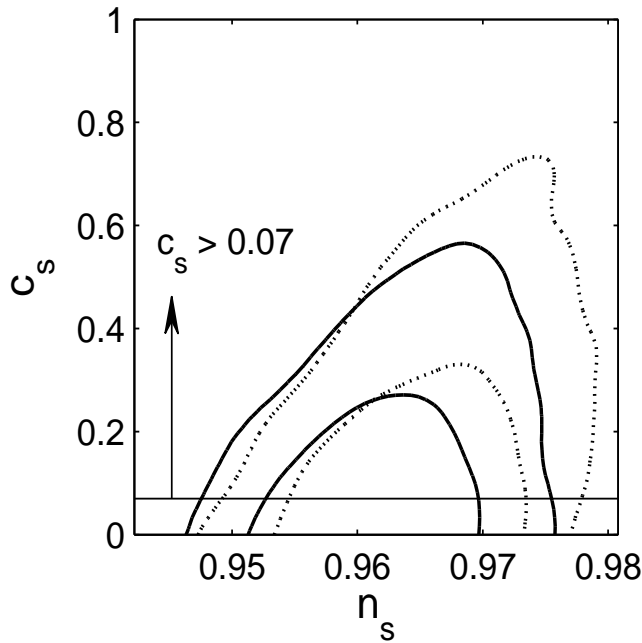}
\caption{\label{fig8}
2-dimensional observational constraints in the $(n_s,c_s)$ plane
in the DBI model with power-law inflation. 
The bold solid curves show the 68\,\%\,CL (inside) and 
95\,\%\,CL (outside) boundaries derived by 
the joint data analysis of Planck+WP+BAO+high-$\ell$, 
whereas the dotted curves are the 68\,\% and 
95\,\% contours constrained by Planck+WP+BAO. 
We also show the bound $c_s>0.07$ coming from 
the non-Gaussianity.}
\end{figure}
%%%%%%%%%%%%%%%%%%%%%%%%%%%%%%%%%

In Fig.~\ref{fig8}, we plot the 68\,\% and 95\,\%\,CL observational 
contours derived from the joint data analysis of 
Planck+WP+BAO+high-$\ell$ and Planck+WP+BAO.  
{}From the Planck+WP+BAO+high-$\ell$ data,
the scalar propagation speed is constrained to be
\ba
& & 0<c_s<0.17 \qquad (68\,\%\,{\rm CL})\,,
\label{csconDBI1} \\
& & 0<c_s<0.43 \qquad (95\,\%\,{\rm CL})\,.
\label{csconDBI2}
\ea
The upper bounds of $c_s$ come from the fact, for 
$c_s^2$ close to 1, the spectrum approaches the Harrison-Zeldovich one.
If we do not include the high-$\ell$ data, the upper limits
of $c_s$ are slightly weaker.
When $c_s \ll 1$, the observationally allowed values 
of $n_s$ are similar to those shown in Fig.~\ref{fig1} for $r \ll 1$.
In the regime $0<c_s \ll 1$, the scalar spectral index can be 
compatible with the data due to the presence of the exponential potential 
satisfying $c_M c_s \gg 1$.

In the DBI model, the non-linear estimator at the equilateral triangle
is given by $f_{\rm NL}^{\rm eq}=-(35/108)(1/c_s^2-1)$ and 
hence $|f_{\rm NL}^{\rm eq}| \gg 1$ for $c_s \ll 1$. 
The leading-order bispectrum (\ref{bispectrum}) reads
\be
{\cal A}_{\cal R}^{\rm lead} =
-\frac{13\beta }{48}\left( \frac{1}{c_s^2}-1 \right)
S_7^{\rm equil}+\frac{13\beta-14}{48}
\left( \frac{1}{c_s^2}-1 \right)
S_7^{\rm ortho}\,.
\label{bispeDBI}
\ee
In the limit $c_s^2 \ll 1$, we have 
${\cal A}_{\cal R} \simeq 
(-0.324/c_s^2)S_7^{\rm equil}+0.0325S_7^{\rm ortho}$, 
in which case the orthogonal shape provides some 
contribution to the total bispectrum.
The Planck team used the shape function of the DBI model
introduced in Ref.~\cite{DBI2} and obtained the following 
bound \cite{Adenon} 
\be
c_s>0.07 \qquad (95\,\%\,{\rm CL}).
\label{DBIbound}
\ee
Since the bispectrum (\ref{bispeDBI}) is valid for any function of 
$f(\phi)$ and $V(\phi)$ in Eq.~(\ref{DBI}), we can use the constraint
(\ref{DBIbound}) for our power-law DBI model as well.
Combining (\ref{DBIbound}) with the bound (\ref{csconDBI2}), 
the scalar propagation speed is constrained to be 
$0.07<c_s<0.43$ (95\,\%\,CL).

%%%%%%%%%%%%%%%%%%%%
\section{Conclusions}
\label{conclusions}
%%%%%%%%%%%%%%%%%%%%

We have studied observational constraints on single-field inflation 
in the framework of the Horndeski's most general scalar-tensor 
theories. This covers a wide class of gravitational theories 
such as (i) a canonical field with a potential, (ii) a non-minimally coupled
scalar field with the Ricci scalar $R$, (iii) running kinetic couplings
$\omega (\phi)X$, (iv) Brans-Dicke theories (including $f(R)$ gravity), 
(v) potential-driven Galileon inflation, 
(vi) field-derivative couplings $G^{\mu \nu} \partial_{\mu} \phi
\partial_{\nu} \phi$ to the Einstein tensor, and (vii) k-inflation.
Under the slow-variation approximation the inflationary observables
like $n_s$, $r$, $n_t$, and $f_{\rm NL}$ can be evaluated in a unified 
way for the general action (\ref{action}).

With the recent Planck data, we run the CosmoMC code
by assuming the flat $\Lambda$CDM Universe.
Since the scalar and tensor runnings are of the order of 
$\epsilon^2$ under the slow-variation approximation, 
we set these parameters to be 0 in the likelihood analysis.
The consistency relation between $r (k_0)$ and $n_t (k_0)$ is different 
depending on the models, so we vary the inflationary observables
${\cal P}_{\cal R} (k_0)$, $n_s (k_0)$, and $r (k_0)$ 
after deriving the consistency relation in each model. 

The difference from the data analysis of the Planck team \cite{Adeinf}
is that we carried out the joint observational constraints with the Planck, 
WP, BAO, and high-$\ell$ data by taking the pivot wave number
$k_0=0.05$~Mpc$^{-1}$ (unlike 
$k_0=0.002$~Mpc$^{-1}$ used by the Planck team).
We confirmed that the joint analysis of Planck, WP, 
and BAO with the consistency relation $r(k_0)=-8n_t(k_0)$
reproduces the results presented in Ref.~\cite{Adeinf}
very well. By adding the high-$\ell$ data, we find that the upper 
bound of $n_s$ becomes tighter than that derived 
in Fig.~1 of Ref.~\cite{Adeinf}
(which is based on either ``Planck+WP+BAO'' or 
``Planck+WP+high-$\ell$''). 

For each inflationary scenario studied in this paper, 
we summarize the main results as follows.

\begin{itemize}
\item (i) A canonical field with a potential $V(\phi)$

Chaotic inflation with the quadratic potential $V(\phi)=m^2 \phi^2/2$ 
is marginally inside the 95\,\%\,CL region.
{}From the joint data analysis of Planck+WP+BAO+high-$\ell$, 
the potentials $V(\phi)=\lambda \phi^n/n$
with $n=1$ and $n=2/3$ are outside the 95\,\%\,CL
boundary for $N=60$. 
The constraints on the models $n=1$ and $n=2/3$
are tighter than those derived by the Planck team \cite{Adeinf}.

In Natural inflation, the symmetry breaking scale $f$ is 
constrained to be $5.1M_{\rm pl}<f<7.9M_{\rm pl}$ (68\,\%\,CL) 
and $f>4.6 M_{\rm pl}$ (95\,\%\,CL).
The upper bound of $f$ was newly derived.

Hybrid inflation with the potential $V(\phi)=\Lambda^4+m^2 \phi^2/2$
is disfavored from the data in the regime $\Lambda^4 \gg m^2 \phi^2/2$.
In another Hybrid inflation model with the potential 
$V(\phi)=\Lambda^4 [1+c \ln (\phi/\mu)]$, the scalar spectral 
index can be as small as $n_s=0.98$ with a suppressed
tensor-to-scalar ratio, but such a model is outside the 
95\,\%\,CL contour. 
These confirm the results of Ref.~\cite{Adeinf}.

We also studied the potentials of the form $V(\phi)=\Lambda^4 [1-f(\phi)]$, 
where $f(\phi)$ is a function that asymptotically approaches 0 
in the limit $\phi \to \infty$.
For the functions $f(\phi)=e^{-\phi/M}$ and 
$f(\phi)=c_1 \phi^{4/3} e^{-c_2 \phi^{4/3}}$
appearing in D-brane inflation and K\"{a}hler-moduli inflation 
respectively, we have $n_s \simeq 1-2/N$ with a very small 
tensor-to-scalar ratio.
Such models are most favored observationally.

\item (ii) Non-minimally coupled models with $\xi \phi^2 R/2$

For the quadratic potential $V(\phi)=m^2 \phi^2/2$, we derived the new
bound $-4.2 \times 10^{-3}<\xi<-1.1 \times 10^{-3}$ (68\,\%\,CL).
The quartic potential $V(\phi)=\lambda \phi^4/4$ enters the 
68\,\%\,CL region for $\xi<-4.5 \times 10^{-3}$ (which is
consistent with the result of Ref.~\cite{Adeinf}).
In Higgs inflation, we have $n_s \simeq 1-2/N$ and 
$r \simeq 12/N^2$ in the limit $|\xi| \to \infty$, in which 
case the model is well inside the 68\,\%\,CL region.

\item (iii) Running kinetic couplings $\omega(\phi) X$

The coupling $\omega(\phi)=e^{\mu \phi/M_{\rm pl}}$ ($\mu>0$)
can reduce the tensor-to-scalar ratio for the potential 
$V(\phi)=\lambda \phi^n/n$ ($n>0$). In the limit $\mu \gg 1$, 
however, $n_s \simeq 1-1/N=0.983$ for $N=60$, 
in which case the model is outside the 95\,\%\,CL region.
We derived the new bounds $-0.02<\mu<0.57$ (95\,\%\,CL)
for $n=2$ and $0.18<\mu<5.0$ (95\,\%\,CL) for $n=4$.

\item (iv) Brans-Dicke theories

In Brans-Dicke theories, the potential $V(\phi)=(3M^2/4) (\phi-M_{\rm pl})^2$
reproduces the Starobinsky's model $f(R)=R+R^2/(6M^2)$ for 
$\omega_{\rm BD}=0$. We test inflation in BD theories for the same 
potential with arbitrary BD parameters $\omega_{\rm BD}$. 
For $|\omega_{\rm BD}| \lesssim 1$, we have 
$n_s \simeq 1-2/N$ and $r \simeq 4(3+2\omega_{\rm BD})/N^2$.
The BD parameter is constrained to be 
$\omega_{\rm BD}<11.5$ (68\,\%\,CL).

\item (v) Potential-driven Galileon inflation

With the Galileon self-interactions $G_3=-X/M^3$, 
$G_4=X^2/M^6$, $G_5=-3X^2/M^9$, the tensor-to-scalar ratio 
can get smaller relative to the case of standard potential-driven inflation. 
For the quadratic potential $V(\phi)=m^2 \phi^2/2$,
the model is inside the 95\,\%\,CL contour, but
outside the 68\,\%\,CL region for $N=60$. 
The quartic potential $V(\phi)=\lambda \phi^4/4$ with
$G_3=-X/M^3$ is outside the 95\,\%\,CL region for $N=60$.
In the presence of the terms $G_4=X^2/M^6$ or $G_5=-3X^2/M^9$, 
however, there are some parameter spaces in which the quartic 
potential is inside the 95\,\%\,CL boundary.

\item (vi) Field-derivative couplings $G^{\mu \nu} \partial_{\mu} \phi
\partial_{\nu} \phi/(2M^2)$

In the presence of the field-derivative couplings to the Einstein tensor, 
the tensor-to-scalar ratio reduces as in the case of potential-driven 
Galileon inflation. However, the potentials $V(\phi)=\lambda \phi^n/n$ 
with $n=2$ and $n=4$ are outside the 68\,\%\,CL region.
We find that the parameter $\alpha=\lambda M_{\rm pl}^{n-2}/M^2$
is constrained to be $\alpha<0.3$ for $n=2$ and 
$\alpha>9.0 \times 10^{-5}$
for $n=4$ at 95\,\%\,CL.

\item (vii) K-inflation

We studied the power-law k-inflation scenario described by 
the Lagrangian (\ref{scalinglag}), in which case $c_s$ is constant.
In the dilatonic ghost condensate model, the observables $n_s$, 
$n_t$, and $r$ are expressed in terms of $c_s$ alone.
{}From the joint data analysis of Planck+WP+BAO+high-$\ell$,  
we derived the bound $0.034<c_s<0.046$ (95\,\%\,CL).
This is not compatible with the bound $c_s>0.079$ (95\,\%\,CL)
coming from the non-Gaussianity with an equilateral template.
In the DBI model with power-law inflation, there are two model 
parameters $c_s$ and $c_M$ in the expressions of 
$n_s$, $n_t$, and $r$.
In this case, the scalar propagation speed is constrained to be 
$0<c_s<0.43$ (95\,\%\,CL). 
Since the bound of $c_s$ coming from the non-Gaussianity is 
$c_s>0.07$ (95\,\%\,CL), there are some allowed parameter 
spaces compatible with both constraints.

\end{itemize}

The above results show that the models with $n_s \simeq 1-2/N$
and the suppressed tensor-to-scalar ratio are most favored observationally.
These include D-brane/K\"{a}hler-moduli inflation, non-minimally 
coupled Higgs inflation, and Brans-Dicke theory in the presence 
of the potential $V(\phi)=(3M^2/4) (\phi-M_{\rm pl})^2$
with the BD parameter $\omega_{\rm BD} <{\cal O}(1)$.
In natural inflation and non-minimally coupled models with 
the potential $V(\phi)=m^2 \phi^2/2$, there are also some allowed 
parameter spaces inside the 68\,\%\,CL region.
Most of other models studied in this paper are outside 
the 68\,\%\,CL boundary.

Although we have focused on the single-field models
in the framework of Horndeski's theories, there are other 
single-field inflationary scenarios based on braneworld \cite{braneworld}, 
non-commutative space-time \cite{noncom}, and 
loop quantum gravity \cite{loop}.
We leave observational constraints on those models 
for future work. 

%%%%%%%%%%%%%%%%%%%%%%%%%%%%%%%%
\section*{ACKNOWLEDGEMENTS}
We thank Nicola Bartolo for useful correspondence.
J.~O. and S.~T. are supported by the Scientific Research Fund of the 
JSPS (Nos.~23\,$\cdot$\,6781 and 24540286). 
S.~T. also thanks financial support from Scientific Research 
on Innovative Areas (No.~21111006).  S. K. is supported
 by the Grant-in-Aid for Scientific research No. 24740149.
%%%%%%%%%%%%%%%%%%%%%%%%%%%%%%%%

%%%%%%%%%%%%%%%%

\end{document}